\documentclass[letterpaper]{JHEP3}
\pdfoutput=1

\usepackage{amsmath}
\usepackage{amssymb}
\usepackage{amsfonts}
\usepackage{lscape, graphicx}
\usepackage{cite}       
\usepackage{bm}         
\usepackage{verbatim}
\newcommand{\iless}[1]{\bigg\lfloor #1 \bigg\rfloor}

\usepackage[all]{xy}
\advance\parskip 1.0pt plus 1.0pt minus 2.0pt   
\def\href#1#2{#2}	

\def\coeff#1#2{{\textstyle {\frac {#1}{#2}}}}
\def\half{\coeff 12}

\def\R{{\mathbb R}}

\def\tr{{\rm tr}}

\def\Z{{\mathbb Z}}

\def\Dslash{{\rlap{\raise 1pt \hbox{$\>/$}}D}}
\def\O{{\cal O}}

\preprint{SLAC-PUB-13798}

\title{ Conformality or  confinement (II):   \\ One-flavor CFTs and mixed-representation QCD
}

\author
    {
    {
    \def\href#1#2{#2}	
    Erich Poppitz$^1$\footnote{\email{poppitz@physics.utoronto.ca}}~
    and Mithat \"Unsal$^2$\footnote{\email{unsal@slac.stanford.edu}}~
           \\${}^1$Department of Physics, University of Toronto,
    Toronto, ON M5S 1A7, Canada
     \\${}^2$SLAC and Physics Department, Stanford University, Stanford, CA 94025/94305, USA
        }
    }%

\abstract{

\smallskip

{\small{
 We study QCD-like four dimensional  theories in the 
theoretically controlled framework of deformation theory and/or twisted
partition function on $ S^1 \times {\mathbb R}^{3}$. By using duality,
we show that  a class of one-flavor theories exhibit new physical
phenomena:  discrete  chiral symmetry breaking  ($\chi$SB)
 induced by the  condensation of topological disorder
operators,  and confinement and the generation of mass gap  due to  new non-selfdual   topological
excitations. In the $\R^4$ limit,  we argue that the  mass gap
disappears,  the $\chi$SB vacua are of runaway type, and  
the theory flows to a  CFT. We  also study  mixed-representation
theories and  find   abelian $\chi$SB  by topological  operators
charged under abelian chiral symmetries. These are reminiscent to,  but
distinct, from  Seiberg-Witten theory with matter, where 4d monopoles
have non-abelian chiral charge. This examination  also helps us   refine
our recent bounds  on the conformal window. In an Addendum, we also discuss mixed vectorlike/chiral representation theories, obtain bounds on their conformal windows, and  compare with the all-order beta function results of arXiv:0911.0931.
 
   }}}

\begin{document}

\maketitle
 
 \section{Conformality or confinement: Introduction and summary}

A new method  to determine the long distance behavior of  asymptotically-free nonabelian gauge theories  with fermionic matter was presented in  \cite{Poppitz:2009uq}. The basic idea is to employ  
 the mass gap for gauge fluctuations as an  invariant characterization of  conformality 
 versus  confinement.\footnote{The notion of the mass gap of a theory is different from what we call the mass gap for gauge fluctuations; see Appendices A, B for discussion.}
 Deformation theory and/or the twisted partition function  permit a controlled calculation of the mass gap for gauge fluctuations in the theory compactified on $S^1 \times \R^3$  for a finite 
 (or sometimes infinite) domain of $S^1$ sizes \cite{Unsal:2007jx,Shifman:2008ja,Unsal:2008ch}. 
 The deformations  were also independently proposed as a useful tool to study phases with partial center symmetry breaking on the lattice \cite{Myers:2007vc, Myers:2009df, Meisinger:2009ne}. 
  The idea to use the mass gap as a characterization of the conformal window has  also been exploited in the worldline formalism  \cite{Armoni:2009jn}. The analysis of \cite{Poppitz:2009uq} was generalized to all classical Lie groups in \cite{Golkar:2009aq}.
 
\begin{figure}[h]
\centering
\includegraphics[width=4in]{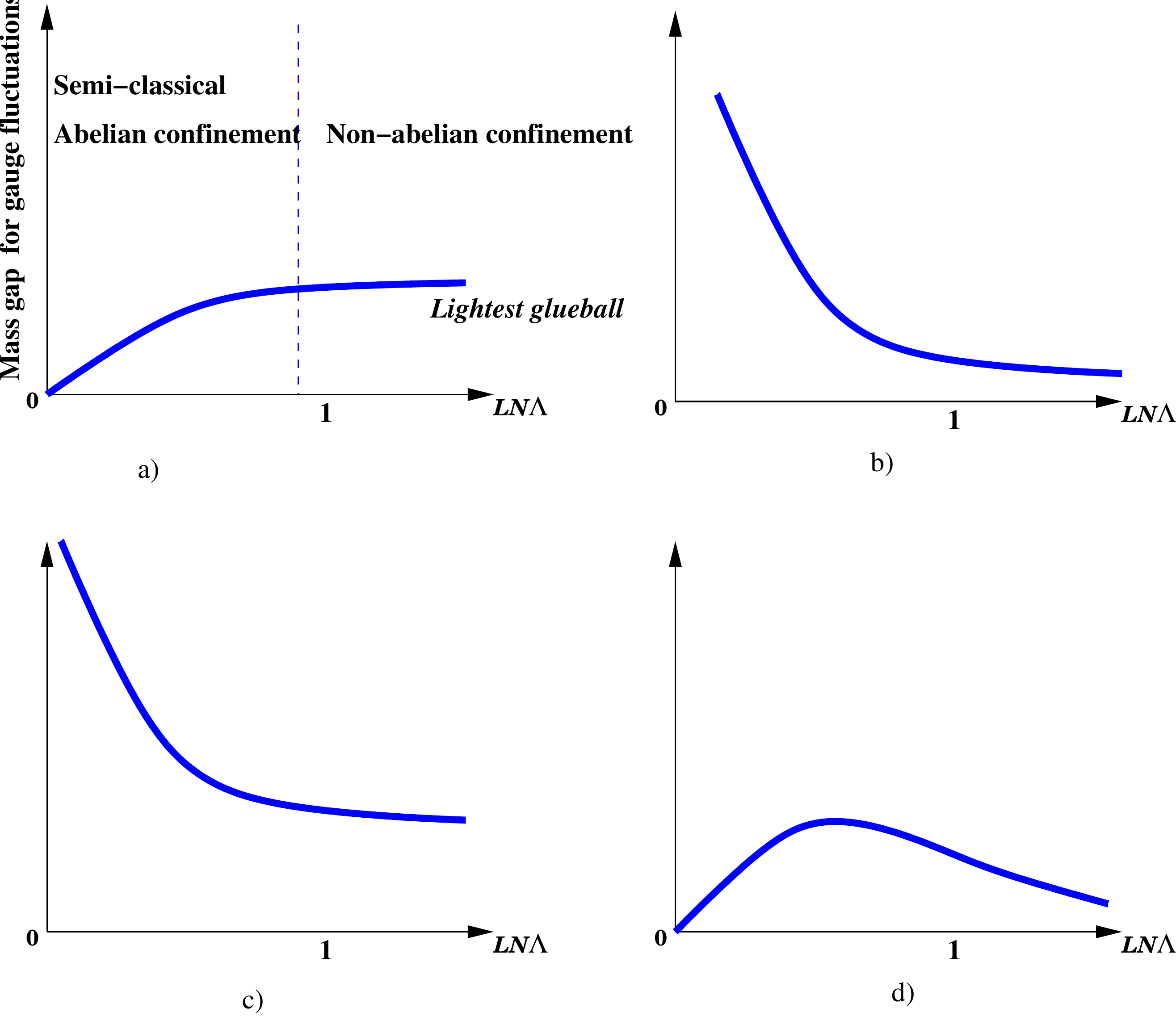}
\caption{Possible behavior of the mass gap for gauge fluctuations in asymptotically free, center-symmetric theories as a function of the size $L$ of ${S}^1$. The semiclassical analysis is valid at $L N \Lambda\ll 1$, where $N$ is the number of colors, and $\Lambda$ is either the strong scale, for confining theories, or the scale where the running coupling  saturates its IR fixed-point value.}
 \label{fig:massgap}
 \end{figure}

 In     \cite{Poppitz:2009uq}, we partially showed and in part conjectured that   gauge theories  fall in one of  four classes---that we refer to as class-${\bf a}$, -${\bf b}$, -${\bf c}$, or -${\bf d}$---with respect to the behavior of the mass gap for gauge fluctuations as a function of the $S^1$ radius $L$. The four possible behaviors, also shown on Fig.~\ref{fig:massgap}, are:
\begin{itemize}
\item The behavior of Fig.~\ref{fig:massgap}{\bf a.)}  holds for  a small  or vanishing number, $N_f$, of massless fermionic species (``flavors" in QCD-like theories). One can show,  in the controlled semiclassical domain of abelian confinement,   that the  mass gap for gauge fluctuations increases with the radius of $S^1$  and  
conjecture that it saturates to its ${\R}^4$  value in the non-abelian   strongly-coupled  confinement domain. 
\item The behavior of Fig.~\ref{fig:massgap}{\bf b.)} holds  for $N_f$  sufficiently large, perhaps  just below the asymptotic freedom boundary on the number of fermions, $N_f^{AF}$. The  mass gap is a decreasing function of the radius at small $S^1$ and decreases further upon approaching $\R^4$. There are theories in this class for which the semiclassical analysis applies at any size  $S^1$. 
\item  Fig.~\ref{fig:massgap}{\bf c.)}  shows a mass gap that  decreases with radius in the semiclassical domain but then saturates to a finite value  on  ${\R}^4$. This can happen, for example, if $\chi$SB takes place on the way.  
\item  Fig.~\ref{fig:massgap}{\bf d.)} shows a mass gap starting to increase with the radius in the semiclassical domain, however, before reaching $\Lambda N L \sim 1$, the coupling reaches a fixed point value without triggering $\chi$SB and the mass gap decreases to zero on $\R^4$. 
\end{itemize}

In \cite{Poppitz:2009uq}, we argued   that small- and large-$N_f$ theories  are class-{\bf a} and class-{\bf b}, respectively, in all of the theories considered. The value of $N_f$ where the small-$LN\Lambda$ behavior of the mass gap switches from increasing to decreasing
with $L$ was taken as an estimate of the lower boundary of the conformal window, $N_f^*$.
However,  apart from admitting the  logical possibility, we did not show whether class-{\bf c} or class-{\bf d} behavior occurs in any gauge theory.\footnote{Except the class-{\bf c}  window between the estimate of  \cite{Poppitz:2009uq}, $N_f^* = 2.5 N$, and   $N_f = 2.61 N$, where the sign of the second coefficient of the beta function changes.}  

\subsection{Single-flavor CFTs}

In this paper, we will first provide an example of a class-{\bf d} gauge theory. It turns out that  this behavior is possible  for a very interesting class of  gauge theories, which flow  to CFTs in the 
IR.   In order for class-${\bf d}$ behavior to take place, the theory must have an increasing behavior of the mass gap in the very small $S^1$ domain---and thus be a decidedly small-$N_f$ theory---and a weak-coupling fixed point (so that $\chi$SB by a fermion bilinear  is not triggered). 
 Indeed, we will show that there is a somewhat exotic class of one-flavor asymptotically-free gauge theories for which these two behaviors are compatible.  Few examples are  one-flavor $SU(2)$ with 4-index symmetric representation fermions, 
and $SU(3)$ and $SU(4)$  gauge theories with 3-index symmetric representation fermions (for Dirac fermions in the 3-index symmetric-tensor representation  asymptotic freedom is lost at $N=5$). 
 
In Section \ref{solvable}, we study in detail only one example of this class. We will show
  that  the dynamics of a vectorlike $SU(2)$ Yang-Mills theory with a single Weyl fermion with  $j=2$, i.e. the 4-index symmetric-tensor representation, is amenable to an analytical treatment on  arbitrary size  $ S^1 \times {\mathbb R}^{3}$ and   ${\mathbb R}^{4}$.
  At  any finite $ S^1 \times {\mathbb R}^{3}$, it exhibits confinement with discrete $\chi$SB and has isolated vacua. However, the  $\chi$SB happens {\it not} via the condensation of a fermion bilinear but via that of the topological disorder operator, generating a 4d-complex mass gap for fermions. This new and interesting phenomena will be discussed in detail below and  Section~\ref{disorderchisb}. 
  We find bounds on the mass gap,   show that the isolated vacua are of run-away type in the decompactification limit and that the  ${\mathbb R}^{4}$ limit is a CFT. 
  To the best of our knowledge, 
among theories without continuous global symmetries, this is the first example of a theory that flows to an interacting CFT.
  
\subsection{Mixed-representation QCD and improvement of the conformality bound}

Our next goal is to determine the theories that may be class-{\bf c}. Especially for  theories  with fundamental fermions, we suspect that there are many theories in this class and that the 
conformal window starts at a value larger  than the critical $N_{\rm F}^* =2.61 N$   obtained in  \cite{Poppitz:2009uq}.  The theories with $N_{\rm F} < 2.5 N$ are   class-{\bf a}  and we want to know which theories are   class-{\bf c} and -{\bf d}.

In Section~\ref{mixed}, we study this question using the following  strategy.  Consider a mixed-representation QCD with $N_{\rm adj}$ adjoint Weyl and $N_{\rm F}$ fundamental Dirac fermions, which we call QCD(adj/F).
Note that, for perturbative purposes, one adjoint Weyl is approximately $N$ Dirac fundamental fermions. As in \cite{Poppitz:2009uq}, we calculate the mass gap for gauge fluctuations in  QCD(adj/F) in the semiclassical domain and obtain:
\begin{eqnarray}
  \label{bion1}
m_\sigma  \sim  \Lambda(\Lambda L)^{\frac{4}{3}( 4-N_{\rm adj} - 
 \frac{N_{\rm F}}{N} )},  \ \    {\rm for}   \;  N_{\rm adj} \geq 1~,
  \end{eqnarray}
  while the result of  \cite{Poppitz:2009uq} for the $N_{\rm adj}=0$ theory is:
  \begin{eqnarray}
  \label{mono1}
m_\sigma  \sim    \Lambda(\Lambda L)^{ \frac{1}{3}( \frac{5}{2}- 
 \frac{N_{\rm F}}{N} )}.
\end{eqnarray}
In theories with  $N_{\rm adj}$$\geq$$1$, the mass gap $m_\sigma$ in   (\ref{bion1})  is induced by  magnetic bions, as  dictated by the relevant index theorem \cite{Nye:2000eg, Poppitz:2008hr}, while at $N_{\rm adj}$$=$$0$, $m_\sigma$ in (\ref{mono1})   is due to monopoles (this explains  why    (\ref{bion1}) does {\it not} reduce to (\ref{mono1})  when $N_{\rm adj}$$=$$0$). 
 
As eqn.~(\ref{bion1}) indicates, once we take $N_{\rm adj}$$=$$1$,  the critical number of fundamental fermions---at which the mass gap changes from an increasing to a decreasing function as a function of $L$ at fixed $\Lambda$---increases to $N_{\rm F}$$=$$3N$. Since adding more massless fermions increases their screening effect, one does not expect the size of the conformal window to decrease. Thus, we take the increase of the critical number of fundamental fermions upon adding a single adjoint as an indication that  some QCD(F) theories with $N_{\rm F} > 2.5 N$ are class-{\bf c}. 

Now, in an attempt to estimate where the class-{\bf c} behavior ends, 
 we trade  the adjoint with $N$ fundamentals.  If  QCD(adj/F) theories with  ($N_{\rm adj}$=$1$, $N_{\rm F}$$<$$3N$) are class-{\bf a}, then 
 QCD(F) theories with    $2.5N$$<$$N_{\rm F}$$<$$4N$ belong to
class-{\bf c}. However, it is also possible that 
some QCD(adj/F) theories in the same domain, but with $N_{\rm F} $ 
close to  $3N$, exhibit   type-{\bf d} behavior. In that case, their ``image" QCD(F) theories are expected to exhibit class-{\bf b} behavior, instead of  {\bf c}, since for such values of $N_{\rm F}$ the mass gap at small $L$ decreases with radius. Thus, we expect  $N_{\rm F} = 4N$ 
to be an upper bound for the lower boundary of conformal window for QCD(F).

The so ``refined" upper bounds on the lower boundary of the conformal window in QCD-like theories with Dirac flavors in the fundamental, symmetric, antisymmetric, and adjoint representations are shown on Fig.~\ref{improvefigure}. We note that for two-index representations the ``refined" and ``un-refined" older estimates of \cite{Poppitz:2009uq} coincide. This is because for two-index representations confinement is predominantly due to magnetic bions, see \cite{Poppitz:2009uq}, thus the analogue of the different behaviors (\ref{bion1}) and (\ref{mono1}) for QCD(adj/F) do not appear upon replacing fundamentals with a two-index representation flavors.  We note that our ``refined"  estimates come remarkably close to those of other older or recent analytical approaches, which are referred 
to in Section~\ref{refinement}.

\subsection{Chiral symmetry and disorder operators}

The  formalism of refs.~\cite{Unsal:2007jx,Shifman:2008ja}  revealed  the existence  of a large class of new non-selfdual topological excitations, which are responsible for confinement on  $S^1 \times \R^3$.  
In Section~\ref{disorderchisb}, we discuss several new phenomena tied with chiral symmetry 
that we observed in our analysis of the single-flavor theories of 
Section~\ref{solvable} and the  mixed-representation ones of Section~\ref{mixed}. In both classes of theories,  as a direct consequence of the index theorem \cite{Nye:2000eg, Poppitz:2008hr},  the pure flux parts of the  monopole-instanton induced  operators  (by ``pure" we mean the operators stripped from their fermionic  zero-modes) are charged under the anomaly-free abelian discrete or continuous global chiral symmetries of the theory, but not under its nonabelian chiral symmetries. This is interesting, because it tells us that  apart from local 
fermion bilinears, there also exist topological disorder operators  
(introduced  by 't Hooft  \cite{'tHooft:1977hy}) charged under abelian chiral symmetries. We show that  these  discrete or continuous anomaly-free abelian chiral symmetries are  spontaneously broken by  the expectation value of these  topological disorder operators---and not by fermion bilinears acquiring a vev---in the small-$S^1$ domain. 

The combination of abelian duality \cite{Polyakov:1976fu} and index theorem  \cite{Nye:2000eg, Poppitz:2008hr}  in the long-distance theories on 
small $S^1 \times \R^3$ map the  dynamical $\chi$SB (abelian) into spontaneous breaking by a tree-level potential, itself induced by non-selfdual topological excitations. This mechanism of $\chi$SB  is similar, but not identical, to Seiberg-Witten theory with matter---where 4d monopole particles carrying nonabelian chiral charge can condense and break chiral symmetry upon the addition of an ${\cal{N}}=1$ supersymmetric deformation \cite{Seiberg:1994aj}.  Up to our knowledge, the realization that topological disorder operators may lead to $\chi$SB is a new phenomenon in QCD-like gauge theories.

The breaking of the abelian chiral symmetry on $S^1 \times \R^3$ occurs already in the weak-coupling limit where the anomalous dimension of various fermion-bilinear order parameters $\gamma_{\bar\psi \psi} (L) \ll 1$, hence  summing ladder diagrams would not lead to $\chi$SB in the gap equation. This is a shortcoming of (resummed) perturbation theory. 

Interestingly, there is also a relation between the anomalous dimension of the  fermion bilinear and 
the action of  a fundamental monopole  $S_0$---which is equal to $1/N$-th of the 4d instanton action and hence survives a large-$N$ limit---which may provide some insight to why 
the ladder approximation and our formalism produce very close estimates.  At one-loop level in perturbation theory and  leading order in the semiclassical  
expansion:
\begin{equation}
\gamma_{\bar\psi \psi}(L) =  {3 \over 2}{ N g^2(L) \over 8 \pi^2}, \qquad 
S_0(L) =    {8 \pi^2 \over N g^2(L)}, \qquad  \gamma_{\bar\psi \psi} \times S_0    =    {\cal{O}}(1)=   {\cal{O}}(N^{0})~,
\label{curious} 
\end{equation}
where $\gamma$ is given for the fundamental of $SU(N)$  at large $N$ (it equals twice that value for two-index representations).   
 At small-$L$, eqn.~(\ref{curious}) is obeyed with  $S_0 \gg 1$ and $\gamma_{\bar\psi \psi} \ll 1$, justifying the use of perturbation theory and the semiclassical expansion. 
Thus,   in the small-$S^1$ domain, only the abelian chiral symmetries are broken by the mechanism described above.
Upon increasing $L$, there are two possibilities: 
\begin{enumerate}
\item 
If $\gamma(L) $  can ever reach unity, this also implies that monopoles and bions will reach a non-dilute regime.  In this case, the semiclassical approximation will break down at 
$LN\Lambda \sim 1$. Beyond $LN\Lambda \sim 1$ is the domain of nonabelian confinement. 
\item 
If $\gamma(L) $   can  never reach to one and remains small at any radius, monopoles and bions will remain dilute at any radius. 
In this case, semi-classical analysis 
may be valid at all radii, and the theories exhibit abelian confinement at finite $S^1$. 
\end{enumerate}
In the first case,  multifermion operators generated by monopoles become strong and are expected to trigger both abelian and nonabelian $\chi$SB.
   This is similar to 
the  $\chi$SB in  the Nambu-Jona-Lasinio (NJL) model. A simple  analysis demonstrates that this phenomenon occurs at $LN\Lambda \sim 1$, the boundary of reliability of the semiclassical analysis. In the second case, multi-fermion operators can never become sufficiently strong to induce  non-abelian $\chi$SB. 
 We suspect that the  curious relation  (\ref{curious}) is the reason why the unrelated 
ladder approximation (based on perturbation theory) and mass gap criterion of deformation theory 
 (based on topological excitations)  produce such close estimates for the conformal window boundary. 

To sum up, our formalism provides a derivation of   confinement  and abelian  $\chi$SB 
within the domain where the semiclassical analysis is reliable and gives some insight regarding the mechanisms of confinement and $\chi$SB.\footnote{
A pertinent  outcome of this analysis, 
 which goes against the common lore,  as reviewed in \cite{Poppitz:2009uq}, is that the mechanism of confinement  in a given gauge theory is dependent on the representation of fermionic matter, in a way dictated by  the index theorems  \cite{Nye:2000eg, Poppitz:2008hr}.}
If pushed to the boundary of its region of validity, it also accommodates non-abelian $\chi$SB 
by naturally generating a sufficiently strong NJL-model, for the first class of theories described above. 

\section{Solvable one-flavor QCD-like theories and  a new class  of CFTs}
\label{solvable}

The standard expectation regarding one-flavor QCD-like theories is confinement, mass gap, and (discrete) chiral symmetry breaking.
In this Section, we argue for the existence of   a new class of QCD-like CFTs within the 
world of one-flavor theories. Moreover, these theories are semiclassically solvable on $S^1 \times \R^3$ of any radius.

Although  our idea is 
inspired by the Banks-Zaks (BZ) limit \cite{Banks:1981nn}, it is  also opposite to it in some sense.  In BZ ($SU(N)$ QCD with $N_f$ fundamental Dirac flavors), one takes the large-$N$, large-$N_f$ limit and dials  $\frac{N_f}{5.5N} = 1 - \epsilon$, where $\epsilon \ll 1$.  
We propose, instead, to  consider an $N_f=1$ theory and use the number of indices in the representation of fermion fields as a parameter. 
For example, for one 3-index symmetric  representation Dirac fermion, $N=2, 3, 4$  theories have asymptotic freedom, while for the 4-index symmetric representation,  only $SU(2)$  with one-Weyl fermion is asymptotically free. In both cases, 
 the two-loop  perturbative $\beta$-function has a fixed point at weak coupling, as weak as, for example, $N_F$$=$$15$  $SU(3)$ QCD. However, unlike the BZ limit, this is not a tunable  coupling.\footnote{In 3d, a  tunably small fixed-point coupling can be achieved by using multi-index representations \cite{Poppitz:2009kz}.} 
 
A  topological distinction  is manifest in $SU(2)$ gauge theories  with Weyl fermions in a spin-$j$ representation. Theories for which $2j=1 \rm (mod 4)$ do  not exist, due to the global Witten anomaly.  The theories with  $2j=1,3 \rm (mod 4)$ are chiral, and  the ones with $2j=0,2 \rm (mod 4)$ are vectorlike. A fermion bilinear does not exist (or vanishes identically) in the  chiral case, and is nonvanishing for the vectorlike theories, for example:
\begin{eqnarray}
&& \psi^2 =  \epsilon^{\alpha_1 \alpha_2} \epsilon^{a_1 a_2}  \psi_{ \alpha_1, a_1 }  \psi_{\alpha_2, a_2} =0  ,\qquad j=\frac{1}{2} ,\cr 
&& \psi^2 =  
\epsilon^{\alpha_1 \alpha_2} \epsilon^{a_1 a_2}  \epsilon^{b_1 b_2}  
\psi_{ \alpha_1, a_1 b_1 }  \psi_{\alpha_2, a_2 b_2 }  \neq  0 , \qquad j=1,
\end{eqnarray} 
where $\alpha$ is the $SL(2,{\mathbb C})$ index for a Weyl fermion. The theories for which
   $2j \geq 5$ are infrared free. 
   
   The vectorlike theory with a single Weyl fermion with $j=1$ is ${\cal{N}}=1$ supersymmetric and
    has been the subject of many past studies, while the chiral theory with $j=3/2$ was recently studied in \cite{Poppitz:2009kz} using methods  similar to the ones of this paper. 
   Here, we
consider the $SU(2)$ gauge theory with a single Weyl fermion in the $j=2$  representation.  The   Lagrangian is:
   \begin{equation}
L= \frac{1}{4g^2} \tr  F_{\mu \nu}^2 + \overline \psi i \bar \sigma_{\mu}D_{\mu}  \psi, \qquad  \psi= \psi_{abcd}~.
\end{equation}  
 The classical theory has a chiral $U(1)$  symmetry, $ \psi \rightarrow e^{i \alpha}  \psi$.  However, quantum mechanically, this symmetry reduces to a discrete subgroup due to instanton effects.  A single  instanton has  ${\cal I}_{\rm inst}= 20$ zero modes and the 't Hooft interaction:
  \begin{equation}
 I (x) = e^{ - S_{\rm  inst}} \psi^{20} \label{inst}, \qquad S_{\rm  inst} = \frac{8 \pi^2}{g^2}~,
 \end{equation}
is  only invariant under   ${\mathbb Z}_{20} \subset U(1)$. 
Consequently,   the chiral symmetry of the quantum theory  is:
 \begin{equation}
 {\mathbb Z}_{20}: \psi \rightarrow e^{i \frac{2 \pi k}{20}   }\psi,  \qquad k=1, \ldots   20.
 \label{Z20}
\end{equation}
A ${\mathbb Z}_{2}$ subgroup of ${\mathbb Z}_{20}$ is fermion number modulo two 
 and cannot be spontaneously broken so long as Lorentz symmetry is unbroken.  If chiral symmetry breaks in this theory, the expected  chiral symmetry breaking ($\chi$SB) pattern  is  
${\mathbb Z}_{20} \rightarrow  {\mathbb Z}_{2}$, leading to ten isolated vacua.   
  This theory has two options in the infrared---confinement, mass gap  and discrete $\chi$SB, or 
conformality, absence of mass gap, and absence of $\chi$SB.
  
 In the next few Sections, we will show that most interesting nonperturbative aspects of this theory can be derived on 
   $S^1 \times \R^3$ geometry for any value of $0<L<\infty$. We show that this theory exhibits 
   mass gap  and discrete $\chi$SB in the $0<L<\infty$ range and has ten isolated vacua. However,  in the $L\rightarrow \infty$ limit, the mass gap vanishes and the ten isolated vacua run away to infinity. 
 We demonstrate that the results on finite but arbitrary   $S^1 \times \R^3$ can be used 
  to put a rigorous bound on the mass gap of the theory on $\R^4$ and claim that the theory on $\R^4$ is a (not too strongly) interacting  CFT.

\subsection{Twisted partition function, Wilson-line eigenvalues, and Buridan's donkey }
    \label{twistedwilson}
    
     We use the  twisted partition function to study the dynamics of this class of theories 
   on  ${\mathbb R}^3 \times S^1$:
\begin{equation}
\widetilde Z(L)= \tr \left[ e^{-L H} (-1)^F \right]~.
\label{tpf}
\end{equation}
Let $\Omega(x)= e^{i \int A_4(x,x_4) dx_4}$ denote the holonomy along the compact direction. It can be brought into a diagonal gauge, shown below for the fundamental representation:
\begin{equation}
\label{vevatinfinity}
\Omega(x \rightarrow \infty)= \left( \begin{array}{cc}
 e^{i v} & 0 \\ 
 0 & e^{-iv} 
 \end{array} 
 \right)~.
\end{equation}
In an appropriate range of  $L$,  where the gauge coupling is small, 
we may evaluate the one-loop effective potential for the holonomy at infinity, $\Omega$, reliably.   The result is: 
\begin{equation}
V_{\cal R}^{+}[\Omega] = \frac{2}{\pi^2 L^4} \sum_{n=1}^{\infty} \frac{1}{n^4} \left[  - \tr_{\rm adj} \Omega^n  +  
 \tr_{\cal R} \Omega^n   \right]~.
 \label{pot1}
 \end{equation}
 The first term is due to  gauge fluctuations and the second term is induced by 
Weyl fermions in a representation ${\cal R}$ endowed with periodic boundary condition as per 
(\ref{tpf}).
 There are $\O(g^2)$ corrections to this formula, which are negligible so long as the running coupling constant remains small.  
 Using character formulas relating the trace in a spin-$j$ representation to that in the defining representation, denoted simply by ${\rm tr}$:
\begin{eqnarray}
\tr_{j=1}(\Omega)=
  (\tr \Omega)^2 -1, \qquad   
\tr_{j=2}(\Omega)
=  
 (\tr \Omega)^4   - 3 ( \tr \Omega)^2 + 1 \, ,
 \label{id2}
\end{eqnarray}
we obtain: 
 \begin{eqnarray}
 \label{oneloop}
   V_{4S}^{+}[\Omega] &=& \frac{2}{\pi^2 L^4} \sum_{n=1}^{\infty} \frac{1}{n^4} \left[ ( \tr \Omega^n )^4   -   4 (\tr \Omega^n)^2  
   \right] ~.
\end{eqnarray}
This is an interesting  potential. At its minimum, the eigenvalues  are neither coincident nor are they maximally apart---the minimum of our one-loop potential (\ref{oneloop}) is located at 
$ \langle   \frac{\tr \Omega}{N}   \rangle = \pm \frac{1}{\sqrt{2}}$.
This should be compared to the more renowned  values of Wilson-line expectation 
values:
 \begin{equation}
\langle   \frac{\tr \Omega}{N}   \rangle = 1 :\; \; {\rm thermal},  \qquad \langle  \frac{\tr\Omega }{N}   \rangle = 0 :\; \; {\rm center-symmetric}.
 \end{equation}
 For our circle-compactification of the $j=2$ theory, one of the minima is at:  \begin{equation}
\langle \Omega \rangle = \left( \begin{array}{cc}
 e^{i \pi/4} & 0 \\ 
 0 & e^{-i\pi/4} 
 \end{array} 
 \right) , \qquad  \langle   \frac{\tr  \Omega}{N}  \rangle = \frac{1}{\sqrt{2}} \
  \label{vev}
\end{equation}
The physics at the other minimum of (\ref{oneloop}) is identical; an exact $Z_2$ center symmetry in this  theory (matter is in an even ``$N$-ality" representation)  interchanges the two minima, ${\rm tr}\Omega$$\rightarrow$$-{\rm tr} \Omega$. Thus, center   symmetry is in fact broken, but   in  a very unconventional way.  Because  the eigenvalues are well separated,   the 
gauge symmetry is also broken,  $SU(2) \rightarrow U(1)$. The rule of thumb, based on experience with other theories 
in the small-$S^1$ regime is that  if center symmetry is broken, eigenvalues clump  and gauge symmetry remains unbroken. If the center is unbroken, then the eigenvalues repel and gauge symmetry is broken down to the maximal abelian subgroup.  The eigenvalue dynamics of this theory  
fits neither characterization\footnote{Note that center symmetry breaking and confinement are not in conflict here---there is no thermal interpretation of the twisted partition function.} and is  somehow reminiscent of Buridan's donkey.
   
Thus, on $\R^3 \times S^1$,  the gauge structure at  distances larger than the compactification 
scale  reduces to an abelian $U(1)$ gauge theory and   
the theory can be described in terms of the perturbatively massless degrees of freedom. These are the photon (or its dual scalar) and a fermion component left massless by the action of the holonomy vacuum expectation value (vev)  (\ref{vev})  in the four-index symmetric representation. The symmetric $\psi_{abcd}$ can be written as  five-component vector. The action of the background vev on  $\psi_{abcd}$  and the component form of $\psi_{abcd}$ 
 are:
\begin{equation}
\label{psivev}
\langle A_4 L \rangle=  v_0 T_3 \equiv  \left( \begin{array}{ccccc}
  2v_0 &  &&& \\ 
  & v_0 &&& \\
  &&0 && \\
  &&&-v_0 & \\
   &&&&-2v_0
 \end{array} 
 \right)         \qquad  \psi_{abcd} =  \left( \begin{array}{c}
   \psi_{1111}   \\ 
   \psi_{1112}    \\
     \psi_{1122}  \\
    \psi_{1222}  \\
     \psi_{2222} 
 \end{array}  \right) \equiv \left( \begin{array}{c}
   \psi_{+2}   \\ 
   \psi_{+1}    \\
     \psi_{0}  \\
    \psi_{-1}  \\
     \psi_{-2} 
 \end{array} 
 \right) ~.
\end{equation}
This implies  that     $\psi_{1122}$ remains massless  to all orders in perturbation theory. The 
subscript in the  second form of  $\psi_{abcd}$ in (\ref{psivev}) labels the charge under the unbroken $U(1)$, which 
implies that    $\psi_{1122} =\psi_{0}$ is  a non-interacting fermion in the long-distance theory. The charged components of the fermions, on the other hand, acquire 3d real-mass  due to gauge symmetry breaking; note that these mass terms are not 4d Lorentz invariant, but preserve the global chiral symmetry, as it comes from the Dirac operator of the 4d theory. 

This result is valid to all orders in perturbation theory and the IR theory (at distances larger than $L$) is a 3d-Maxwell theory with a  non-interacting fermion:  
 \begin{equation}
{\cal{L}}^{\rm pert. theory}= \frac{L}
{4g^2(L)} F_{3,ij}^2 +  L \overline \psi_0 i \bar \sigma_{i}\partial_{i}  \psi_0 ,
\label{pt} 
\end{equation}  
  We would like to see whether the dual photon and the massless fermion may acquire  mass  
  due to  nonperturbative effects.\footnote{ In the thermal case, the eigenvalues collapse to zero and    there is no length scale at which the $SU(2)$ gauge structure reduces to its Cartan 
subgroup. Moreover,  the fermions  decouple from the IR physics, with a 
 ${\cal{O}}(T)$ thermal mass. This is unlike the spatial compactification, where there are fermionic zero modes surviving in the long distance regime. }

   \subsection{Demonstration  of mass gap}  
   
Since the holonomy (\ref{vevatinfinity}) causing  the gauge symmetry breaking 
$SU(2) \rightarrow U(1)$  is compact, there exist   two types of  monopoles---BPS and Kaluza-Klein (KK)  \cite{Lee:1997vp, Kraan:1998pm}, which we label 
$ {\cal M}_{\rm 1} $ and  ${\cal M}_{\rm 2}$, respectively.
  The actions  $S_{1}$  and $S_{2}$ of these two monopoles are determined by the separation between eigenvalues of the Wilson line.  Their actions relative to the 4d instanton are:
       \begin{eqnarray}
  S_{1} = \frac{1}{4}  S_{\rm inst} , \qquad  S_{2} = \frac{3}{4}  S_{\rm inst} \; ,
  \qquad  
    \end{eqnarray} 
    as opposed to the usual equal actions ($S = S_{\rm inst}/2$) for center-symmetric $SU(2)$ compactifications.
For a background consisting of $n_1$, $n_2$ multiples of these two topological excitations, the  index of the Weyl operator in a spin-$j$ representation  with index $T(j) = (1/3) j(j+1)(2j+1)$ is given  in  \cite{Poppitz:2008hr}:
\begin{equation}
\label{indexstaticSU2any}
{\cal I}_{j}[n_1, n_2] = n_2 \; 2 T(j) - (n_1-n_2)  \sum\limits_{m=-j}^j 2 m \iless{-{m v L \over 2 \pi}},
\end{equation}
 where $v$ is the expectation value of the holonomy (\ref{vevatinfinity}),  $\lfloor x \rfloor$ denotes the largest integer smaller than $x$, and  ${\cal I}_{2}[1,0] = {\cal I}_{\rm BPS} $, 
  ${\cal I}_{2}[0,1] = {\cal I}_{\rm KK} $,  ${\cal I}_{2}[1,1] = {\cal I}_{\rm inst.} $.  
  Thus, for $j=2$, the indices for the the BPS, KK, and 4d instanton are:\footnote{It would be interesting to see how these  topological excitations and their indices arise by studying  orthogonal combinations of calorons \cite{Kraan:1998pm}  along the lines of \cite{GarciaPerez:2009mg}.}
\begin{equation}
{\cal I}_{\rm BPS} =6, \qquad   {\cal I}_{\rm KK} =14, \qquad    {\cal I}_{\rm inst}= {\cal I}_{\rm BPS} + {\cal I}_{\rm KK} =20.
\end{equation}
This implies that the leading    (anti)monopole-induced operators are:
   \begin{eqnarray}
 &&  {\cal M}_{\rm 1} = e^{-S_{1}} e^{i \sigma} \psi^6, \qquad  \overline {\cal M}_{ 1} = 
   e^{-S_{2}} e^{- i \sigma} {\bar \psi}^6,  \cr \cr 
&&    {\cal M}_{\rm 2} = e^{-S_{1}} e^{- i \sigma} \psi^{14}, \qquad 
     \overline {\cal M}_{2 } =  e^{-S_{2}} e^{ i \sigma} {\bar \psi}^{14} \; .
     \label{mon}
   \end{eqnarray}
   The instanton operator may be viewed as a composite of the two types of monopole operators
   and is given by:
      \begin{eqnarray}
 I \sim  {\cal M}_{\rm 1}    {\cal M}_{\rm 2}  \sim e^{-   S_{\rm inst} }  \; \psi^{20}, \qquad 
 S_{\rm inst} =     S_{1}  + S_{2} = \frac{8 \pi^2}{g^2}~.
    \end{eqnarray}
 
The elementary monopoles (\ref{mon}) and the instanton term (\ref{inst}) do not generate  mass   for the dual photon and   the zero-mode fermions $\psi_0$.   As usual, 
let us first demonstrate on symmetry grounds that a non-perturbative mass  for the photon is allowed. 
Since  ${\mathbb Z}_{20}$   is a non-anomalous symmetry of the microscopic theory, it must also be a symmetry of the long distance theory.  In particular, the invariance of the monopole operator 
$ {\cal M}_{\rm 1}$ 
demands that $e^{i\sigma}$ must transform non-trivially under  ${\mathbb Z}_{20}$: 
\begin{equation}
\begin{array}{c|c}
& \Z_{20}     
 \\ \hline
\psi &  1  \\
\hline \hline
e^{i \sigma}& -6
\label{photonshift}   
\end{array} ~.
\end{equation}
 Thus, the theory in terms of the dual photon possesses a ${\mathbb Z}_{10}$ shift 
 symmetry,\footnote{Clearly, ${\cal{M}}_2$ is also ${\mathbb Z}_{10}$ invariant.} which forbids all purely bosonic operators  $e^{i q \sigma}$  but 
 $q=0$(mod $10$). The leading such purely bosonic operator is: 
 \begin{eqnarray}
  ( e^{i10 \sigma}  +   e^{-i10 \sigma}  )  \sim   \cos 10 \sigma~.
\end{eqnarray}
As strange as it may sound,  this is the first operator which may  generate  a mass gap in the gauge sector, but it is very  suppressed in topological expansion.  Below, we will provide an explanation for this operator in terms of ``elementary"  topological excitations (\ref{mon}). 

The   $e^{i10 \sigma}$ operator can be induced by a  topological excitation with magnetic charge $+ 10$ and zero net index. All fermion zero modes need to be soaked up.  We need 
$x$ $ {\cal M}_{\rm 1}$  with  fermions of one chirality to  be contracted with $(10-x)$ 
$ \overline {\cal M}_{\rm 2}$ which carry fermions of opposite chirality. 
Since the 
  least common multiple of two types of index is: 
\begin{equation}
{\rm l.c.m.}({\cal I}_{\rm BPS}, {\cal I}_{\rm KK})= {\rm l.c.m.}(6,14)= 42,
\end{equation}
it follows that $x=7$. This means that  the  pure flux operator 
   $e^{i10 \sigma}  $   without any fermionic zero modes has 
 the  same quantum numbers as the ten  monopole state with the quantum numbers of   7 BPS and 3 ${\overline {\rm KK}}$ monopoles.  We will refer to this excitation as a {\it magnetic decouplet}. The magnetic decouplet operators are essentially: 
\begin{equation}
{\cal MD}  =  [{\cal M}_{\rm 1}  ]^7    [\overline {\cal M}_{2 }]^3    \sim e^{i 10 \sigma} , 
\qquad \overline {\cal MD}  \sim e^{- i 10 \sigma}~,
\end{equation}
with all fermionic zero modes contracted.  This is the leading object which  generates a mass term.\footnote{The magnetic and topological charges of these excitations are
 $  \left( \int_{S_{ \infty}^2} B,  \int    F \widetilde F  \right) =   \left( \pm 10,    \mp \half \right)   $, 
 where the signs are correlated. } The proliferation of ${\cal MD}$ and $\overline {\cal MD} $, generate, in long-distance effective theory, the potential:
 \begin{equation}
   e^{-7 S_{1} - 3 S_{2}} ( e^{i10 \sigma}  +   e^{-i10 \sigma}  )  \approx    e^{-4 S_{\rm inst}} \cos 10 \sigma
\end{equation}
This term  induces a mass  for the dual photon, as can be seen by expanding the potential around one of its minima. 
Since $\sigma$ is a  variable with period $2 \pi$ and the potential has ten  minima within 
the fundamental domain of $\sigma$, the $ {\mathbb Z}_{10}$  symmetry is spontaneously broken. The ten  minima are located at: 
\begin{equation}
\langle e^{i\sigma} \rangle = e^{i  \frac{2  \pi}  {10} q}, \qquad  q=1, \ldots 10. 
\label{topdis}
 \end{equation} 
 This $ {\mathbb Z}_{10}$, as stated earlier, is the  broken  subgroup of ${\mathbb Z}_{20}$  
 discrete chiral symmetry. 
 
Since the microscopic theory is a one-flavor vector-like theory,   
the discrete $ \chi$SB ${\mathbb Z}_{20} \rightarrow  {\mathbb Z}_{2}$ must generate a 4d complex-mass term for fermions as well. Indeed, another composite, this time of  2 BPS and 1 ${\overline {\rm KK}}$ monopoles has two left-over zero modes after contractions, and is the leading candidate to generate a fermion mass term:  
\begin{equation}
[{\cal M}_{\rm 1}  ]^2    [\overline {\cal M}_{2 }]  + {\rm h.c }  \sim   e^{-2S_{1} - S_{2} } ( e^{i 3\sigma} \overline \psi^2 + e^{-i 3\sigma} \psi^2)  \longrightarrow e^{-\frac{5S_{\rm inst}}{4} }  \; (\overline \psi^2 +  \psi^2 )~.
\end{equation}
In the last stage, we expanded the $\sigma$ field around the minimum at $\sigma=0$. 
In four dimensional one-flavor QCD-like theories, it is expected that  $ \chi$SB will generate a mass gap for fermions. We show that this is indeed the case, however, what is surprising is that 
this phenomena is driven by the condensation of a topological disorder operator (\ref{topdis}).  

The long-distance Lagrangian (\ref{pt}), now corrected for nonperturbative effects and written in terms of the dual photon field $\sigma$, becomes:
\begin{eqnarray}
{\cal L}^{\rm dual}=&& \underbrace{ \frac{g^2}{2L} (\partial \sigma)^2 +  \psi  i \bar \sigma_{i}\partial_{i}  \psi  +  \ldots}_{\rm perturbation \; theory}   
+  \underbrace{ e^{-S_1}  e^{i \sigma}  \psi^6   + \; e^{-S_2}  e^{-i \sigma}  \psi^{14} + {\rm h.c.}    
}_{\rm magnetic\;  monopoles} +  \ldots \cr \cr
+&&  \underbrace{ e^{-2S_{1} - S_{2} } e^{- i 3\sigma}  \psi^2  + {\rm  h.c.} }_{ \rm magnetic \; triplets}
 + \;  \underbrace{  e^{-7S_1 -3S_2} \cos 10 \sigma}_{\rm magnetic \; decouplets}   + \ldots~.
 \label{dual}
 \end{eqnarray}  
In (\ref{dual}), we have normalized the fermion field as appropriate in 3d, omitted powers of $L$ needed to make up the dimensions of the multifermion and potential terms, as well as numerical coefficients and   (uncalculated) powers of $g^2$.
 
 The dual Lagrangian  ${\cal L}^{\rm dual}$ and the physics it encapsulates   are the main results of this Section. 
Ellipsis in (\ref{dual}) represent perturbatively and non-perturbatively generated operators that are subleading at small coupling (certainly, there are also other operators until the tenth order in the semiclassical expansion is reached, but the leading operator generating a mass gap for gauge fluctuations appears only at that order). We also kept the operator which generates the mass gap in the fermionic sector. Expanding the dual Lagrangian to quadratic order  in fields around one of the ten isolated vacua, and restoring the factors of $L$, one obtains 
\begin{eqnarray}
{\cal L}^{\rm dual}_{\rm quadratic}= \frac{g^2}{2L} (\partial \sigma)^2  +  \frac{  1}{L^3} 
e^{-4 S_{\rm inst.}}  \sigma^2  +  \psi  i \bar \sigma_{i}\partial_{i}  \psi  
+   \frac{1}{L} e^{-\frac{5}{4} S_{\rm inst.}}  ( \psi^2  + \overline \psi^2)
\label{dualquad}~.
 \end{eqnarray}

To summarize, we have shown that, at finite $S^1 \times \R^3$ the theory exhibits a mass gap, confinement and  
discrete  $\chi$SB.  If this behavior continues to $\R^4$, this would  imply that this theory is confining on  $\R^4$. 
However,  we will demonstrate below  a rigorous bound on mass gap of gauge fluctuations  valid at any $L$ and show that it vanishes  on decompactification limit, implying conformality on    $\R^4$.

{\flushleft{ \bf Region of validity of the one-loop analysis and the semiclassical expansion:}}
 The range of validity of the one-loop potential 
 depends on whether the gauge coupling is weak or not at the scale of compactification. 
 For confining gauge theories, this implies a small-$L$ domain of validity  $L \ll \Lambda^{-1}$,
  where  $ \Lambda$ is the strong scale of the theory. 
For asymptotically free theories with a weak-coupling IR fixed point, 
 the region of  validity of (\ref{pot1}) extends to all values of $S^1$ radius, i.e.:
 \begin{equation}
g^2 (L)  \leq g^2_*\equiv  g^2(L_*),  \qquad   {\rm  for \; all \; } L  ~.
\end{equation}
In other words, the coupling at the scale of compactification first grows as in any asymptotically-free theory,  for sufficiently small $L$, $L\ll L_*$, and at the scale $L_*$ it saturates to its fixed point value. 
  The fixed point of the two-loop RG beta function is located at:
 \begin{equation}
 g_*^2 = - \frac{16 \pi^2  \beta_0}{\beta_1} \approx 0. 74, \qquad   \frac{g_*^2 }{4 \pi} \ll1~.
 \label{fixed}
 \end{equation}
Thus, the one-loop potential (\ref{pot1}) is reliable\footnote{$g_*^2$ is as weak as the fixed-point coupling of the 15-flavor $SU(3)$ theory,  which is argued to be conformal. However, in the absence of a parametrically tunable fixed point coupling (as opposed to the BZ limit), strictly speaking, we rely on the assumption that higher loops do not introduce large numerical factors. 
\label{concern}
} at any $0< L < \infty$.

\subsection{A non-perturbative  bound on the mass gap and flow to conformality}
\label{boundonmassgap}

First, consider the $L \ll L_*$ domain, where $L_*$ is the saturation scale of  the coupling constant. In this regime, the one-loop result for the  $\beta$ function dominates and  the strong scale is given by:
\begin{eqnarray}
e^{-\frac{8 \pi^2}{g^2(L)}} = (\Lambda L)^{\beta_0}, \qquad \beta_0= \frac{11}{3}
N - \frac{2}{3} T(j) N_f^{W}, \qquad T(j) = \frac{1}{3} j(j+1)(2j+1)\;.
\end{eqnarray}
Setting $j=2$ and  the number of Weyl spinors $N_f^{W}=1$,  we obtain $\beta_0=\frac{2}{3}$.
The mass gap for the dual photon is 
$m_{\sigma} \simeq \frac{1}{L}e^{-2S_{\rm inst}(L)}=  \Lambda(  \Lambda L)^{\frac{1}{3}}$.  
However, for $L> L_{*}$, the coupling constant  reaches its fixed point value. In this domain, 
the mass gap for gauge fluctuations is, up to unimportant prefactors:
\begin{equation}
m (L)  = \left\{ \begin{array}{ll}
 \Lambda(  \Lambda L)^{\frac{1}{3}} & \qquad 0< L < L_{*} \\ 
\frac{1}{L} \;{\rm exp}\left[ {-\frac{16 \pi^2} {g^2_* } } \right]  , & \qquad  L_{*} < L < \infty 
\end{array} 
\right.~.
\label{gap}
\end{equation}

 There is a caveat to the above argument.  If the fixed point is reached in a domain where the monopole operators are non-dilute,   the semiclassical approximation  for $L$ of order and larger than $L_*$  is not to be trusted and along with it,   also the second line in (\ref{gap}). On the other hand, if the topological excitations remain dilute, the result shown in the second line in (\ref{gap}) presents a rigorous bound on the mass gap of the theory, since the gauge coupling at any scale is  smaller or equal to the fixed-point value. 
 
 In the theory at hand, the instanton factor at the fixed point, as well as the monopole fugacities  are actually exponentially  small:  $e^{-S_{\rm inst}}  \sim e^{-\frac{8 \pi^2}{g_*^2} } \sim e^{-106}, \;  e^{-S_{1}}  \sim e^{-\frac{8 \pi^2}{4g_*^2} } \sim e^{-26.75}$. This retrospectively justifies 
 the use of the semiclassical expansion at any value of $0<L<\infty$.  Thus, since $ e^{- \frac{8 \pi^2} {g^2(L)} }  \leq  e^{- \frac{8 \pi^2} {g^2_{*} } }  $, this implies  a tiny upper bound  on the mass gap on 
 $\R^3 \times S^1$ of any size:
 \begin{equation}
 \label{massgaplimit}
m(L)  \leq \frac{1}{L} \exp \left[ {- \frac{16 \pi^2} {g^2_* } } \right] \sim  \frac{1}{L}  e^{-212}. \end{equation}

 Since the dual photon mass\footnote{  
Validity of the effective field theory described by the dual Lagrangian (\ref{dual}) requires a separation of scales between the $W^{\pm}$ boson mass, $\sim 1/L$, and the dual photon. Since $m_\sigma/m_W \sim e^{-{16 \pi^2\over g^2}}$ this separation is manifest at weak coupling. 
Note also that the dual photon is a scalar in the 3d long-distance theory. The small value of $m(L)$ is another example showing how a mass term may in fact be irrelevant and small without fine-tuning, akin to the pseudogoldstone mechanism.}  is approximately equal to 
 $m_{\sigma} \sim \frac{1}{L}e^{-2 S_{\rm inst}} $,  confinement on $\R^3 \times S^1$ sets in 
 at distances  $m_{\sigma}^{-1} \sim L e^{+2 S_{\rm inst}} $ (we also note that in terms of a canonically normalized dual photon field, the $\chi$SB vacua are ``runaway" to infinity in the infinite-$L$ limit).  
We can safely say that it is impossible to see the confining regime of this theory in any practical lattice simulation.\footnote{ If this theory is simulated on a four dimensional asymmetric toroidal lattice with 
 $L_1^3 \times L_2$ sites ($L_1 \gg L_2$),  the low energy limit of this theory will be 
 seen as an abelian Coulomb phase as described in  (\ref{pt}), 
  a  pure non-compact Maxwell theory on $\R^3$ and a massless fermion. 
 If the lattice is toroidal and symmetric, then a dynamical abelianization is not expected to occur.   }
 In the decompactification limit, eqn.~(\ref{massgaplimit}) implies that:
 \begin{equation}
m({\R^4})  = \lim_{L\rightarrow \infty}  m(L) =0~,
 \end{equation}
 showing gaplessness of the theory on $\R^4$. 
  On $\R^4$, we do not expect dynamical abelianization to take place at any length scale in this gauge theory. Rather, we expect, the $W$-boson  components of the gauge fluctuations to remain massless as well. The long distance theory on $\R^4$ is described in terms of short distance quarks and gluons and the long distance lagrangian is the same as the classical lagrangian. 

Thus, the $SU(2)$ theory with a one-flavor four-index symmetric representation Weyl fermion belongs to the 
class of one-flavor CFTs with high-index representations. The results of this Section show that it is an example of class-{\bf d} in our classification shown in Fig.~\ref{fig:massgap}. Other examples are theories with one-flavor 3-index symmetric representations with gauge groups $ SU(3)$ and $ SU(4)$, whose dynamics can be worked out along similar lines (recall that at $N=5$ asymptotic freedom is  lost).

\section{QCD with mixed-representation fermions}
\label{mixed}

Consider now $SU(N)$ Yang-Mills theory with one adjoint Weyl fermion  $\lambda$ and $N_{\rm F}$  Dirac fundamental fermions, $\Psi = \left( \begin{array}{c} \psi_L \cr
\overline \psi_R \end{array}  \right)
$.
 The  global chiral symmetry of the classical theory is
$U(1)_{\lambda} \times U(1)_B  \times U(1)_{\Psi} \times SU(N_{\rm F})_L \times    SU(N_{\rm F})_R$
where the  $U(1)_{\lambda}$ acts on $\lambda$ and  $U(1)_{\Psi}$ is the axial symmetry acting on $\Psi$ in a canonical way. In contradistinction to the theories with one type representation, where there is only one  classical axial symmetry, reduced  to a discrete symmetry by instantons, in theories with mixed representation fermions, instanton effects just reduce the  $U(1)_{\lambda} \times  U(1)_{\Psi}$ to a diagonal axial group $U(1)_A$. 
Inspecting the 't Hooft   vertex: 
\begin{equation}
I(x) =  e^{-S_{\rm inst}} (\lambda \lambda)^{N} \left[ ( \psi_L^1\psi_R^1) \ldots (\psi_L^{N_{\rm F}}\psi_R^{N_{\rm F}}) + \ldots  \right]=   e^{-S_{\rm inst}} (\lambda \lambda)^{N} \det_{I, J} \psi_L^I \psi_R^J \;,
\end{equation}
it is evident that the instanton operator is invariant under the axial $U(1)_A$ generated 
by $Q_A= Q_{\lambda}- \frac{N}{N_{\rm F}} Q_{\Psi}$.
Thus, the  continuous symmetries of the quantum theory are as shown  below (the charges of the various monopole and bion operators are also given in table (\ref{charges0}) and are explained further in this Section):\footnote{Readers familiar 
 with SUSY-QCD will find these transformation properties familiar. In the supersymmetric context,   the chiral $U(1)_{A}$ is an R-symmetry, under which 
the fermions have the same charges. This is, of course, expected because the 
$(N_{\rm adj}, N_{\rm F})=(1, N_{\rm F})$ theory can be obtained by setting the scalar masses in 
SUSY-QCD to infinity. Once this is done, one obtains not the usual QCD(F), but the QCD-like theory with mixed-representation matter.  The chiral symmetries are unaltered by this procedure.}
\begin{equation}
\begin{array}{c|cccc}
& U(1)_B  &  U(1)_{A}  & SU(N_{\rm F})_L &     SU(N_{\rm F})_R  
 \\ \hline
\lambda & 0 & 1& 1& 1  \\
\psi_L &  1&  - \frac{N}{N_{\rm F}} &  \Box & 1 \\
\psi_R &  -1&  - \frac{N}{N_{\rm F}} & 1 &  \Box \\
\hline \hline 
e^{i \alpha_i \sigma } & 0& -2 & 1& 1 \\
e^{i \alpha_N \sigma } & 0& (2N-2)& 1& 1 \\
e^{i (\alpha_i- \alpha_{i+1})  \sigma }  & 0& 0 &1&1\\
e^{i (\alpha_{N-1}- \alpha_{N})  \sigma }  & 0& -2N &1&1 \\
e^{i (\alpha_{N}- \alpha_{1})  \sigma }  & 0& 2N &1&1
\end{array} 
\label{charges0}
\end{equation}
We consider the mixed-representation theory with periodic spin connection for fermions on 
$\R^3 \times S^1$ and apply double-trace deformations  to preserve the center symmetry.\footnote{A generalization of this analysis which incorporates $N_{\rm adj}>1$ Weyl adjoint fermions  does not require double-trace deformations if periodic boundary conditions for the fermions on $S^1$  are used.  For the $N_{\rm adj}=1$ theory, 
the one-loop potential is order $N$ rather than $O(N^2)$. Thus, a tiny deformation is good enough to preserve the (approximate) center symmetry.}

On $\R^4$, the standard expectation regarding this theory, for a low number of fermions,  is spontaneous breaking of all axial symmetries, down to $U(1)_B \times SU(N_{\rm F})_{L+R}$. Note that, one also expects the 
$U(1)_A$, which is an exact symmetry in this theory, to break down spontaneously.  Thus, there must be  $N_{\rm F}^2$ Goldstone bosons, as opposed to   $N_{\rm F}^2-1$. For theories in the conformal window, no symmetry is expected to be broken spontaneously. 
 
 At small $S^1$, there are $N$-types of fundamental BPS and KK monopoles due to gauge symmetry breaking $SU(N) \rightarrow U(1)^{N-1}$ by the Wilson line holonomy. By using the index theorem \cite{Nye:2000eg,Poppitz:2008hr}, we deduce that:\footnote{By using a $U(1)_B$-twist in boundary conditions of fermions, the fundamental zero modes can always be localized to the $N^{\rm th}$ (``Kaluza-Klein") monopole. This is assumed in our charge assignments for monopole operators, is done for convenience, and does not invalidate the generality of the results.}  
\begin{equation}
{\cal M}_i = e^{-S_0} e^{i \alpha_i \sigma } (\alpha_i \lambda)^2, \qquad
 {\cal M}_N = e^{-S_0} e^{i \alpha_N \sigma } (\alpha_N\lambda)^2  \det_{I, J} \psi_L^I \psi_R^J
 \qquad i=1, \ldots N-1 ~,
\end{equation}
where $\alpha_1, \ldots, \alpha_N$ denote the affine roots of $SU(N)$ and $S_0 = {8 \pi^ 2 \over g^2 N}$.
Note that the monopole operators are manifestly invariant under $U(1)_B  \times SU(N_{\rm F})_L \times    SU(N_{\rm F})_R$. Invariance under the anomaly-free axial symmetry $U(1)_A$ demands the charge assignments for the pure 
monopole operators shown in (\ref{charges0}).

Note that the charge assignments are unlike the $N_{\rm F}$$=$$0$ theory (which has ${\cal N}$$=$$1$ supersymmetry), where  all  $N$ monopole operators have charges $-2$ under the anomaly free (discrete) chiral symmetry.    This implies, following the analysis of \cite{Unsal:2007jx} that   there are $N-2$ rather than $N$ magnetic bions where all fermion zero-modes can be soaked-up. 
The two other composites still carry zero modes. The bion operators are:
\begin{eqnarray}
&&{\cal B}_i = {\cal M}_i \overline {\cal M}_{i+1} =   e^{-2S_0} e^{i (\alpha_i  - \alpha_{i+1})
 \sigma } ~,
\qquad    \qquad i=1, \ldots N-2 ~,
\cr \cr
 && {\cal B}_{N-1} =  e^{-2S_0} e^{i (\alpha_{N-1}  - \alpha_{N}) \sigma } 
  \det_{I, J} \overline \psi_L^I  
 \overline\psi_R^J ~, \qquad 
 {\cal B}_{N} = e^{-2S_0} e^{i (\alpha_{N}  - \alpha_{1}) \sigma } 
  \det_{I, J}  \psi_L^I  \psi_R^J  ~.
\end{eqnarray}

Since there are $N-1$ dual photons in the IR, but only $N-2$ of them obtain mass due to the bion-generated potentials,  one dual photon remains massless. 
In order to understand the nature of this photon, it is useful to discuss the simplest example, the 
$SU(2)$  gauge theory with $(N_{\rm adj}, N_{\rm F})=(1,1)$.

\subsection{$\mathbf (N_{\rm adj}$$=$$1,$ $N_{\rm F}$$=$$1$) ${SU(2)}$ gauge theory}
\label{mixed1}
The $SU(2)$ gauge theory with mixed representations hosts some new and interesting phenomena.  On $\R^4$, the instanton operator is:
\begin{equation}
I(x) =  e^{-S_I} (\lambda \lambda)^{2}  ( \psi_L\psi_R) ~,
\end{equation}
and the theory has an exact $U(1)_B \times U(1)_A$ symmetry\footnote{$U(1)_B$ is actually the $T^3$ part of the enhanced $SU(2)$ flavor chiral symmetry acting on $\psi_{L,R}$, which, however, remains unbroken.} with charges: 
\begin{equation}
\begin{array}{c|cc}
& U(1)_B  &  U(1)_{A}   
 \\ \hline
\lambda & 0 & 1  \\
\psi_L &  1&  - 2  \\
\psi_R &  -1&  -2 \\ 
\hline \hline
e^{i \sigma}& 0 & -2
\label{charge}
\end{array} 
\end{equation}
On  $\R^3 \times S^1$, there are two-types of  monopole operators:
\begin{equation}
{\cal M}_1 = e^{-S_0} e^{i  \sigma } \lambda^2, \qquad
 {\cal M}_2 = e^{-S_0} e^{- i  \sigma } \lambda^2   \psi_L \psi_R
\end{equation}
The invariance of monopole operators under $U(1)_A$ demands the dual photon to transform as given in (\ref{charge}).

{\flushleft{\bf Expectations on $\mathbf \R^4$:}} On $\R^4$, this theory is expected to confine and break its chiral 
$U(1)_A$ symmetry.  This implies the existence of one Nambu-Goldstone (NG) boson. The fluctuation around the vacuum can be parameterized as:
\begin{equation}
\langle \lambda \lambda \rangle = \Lambda^3 e^{i \pi/f_{\pi}}, \qquad \langle  \psi_L \psi_R \rangle = \Lambda^3 e^{-2 i \pi/f_\pi}
\label{chiralc}
\end{equation}
where $\pi$ is the massless ``pion." It is also expected that for $L > \Lambda^{-1}$, a long distance description based on the ``chiral Lagrangian" of the pion will be adequate. 

{\flushleft{\bf Center-stabilized theory on small $\mathbf S^1 \times \R^3$:}}  Because of gauge symmetry breaking $SU(2) \rightarrow U(1)$, the long distance theory can be described in terms of  the photon and  the component of the adjoint fermion along the Cartan subalgebra. The center symmetric vev $\langle A_4L \rangle = {\rm Diag}\left(\frac{\pi}{2}, -\frac{\pi}{2} \right)$ generates a 3d-real mass term for the fundamental fermions. The real mass term, as opposed to 4d-complex mass term, respects chiral symmetry. The effective Lagrangian for the perturbatively massless modes is denoted by ${\cal L}^{0}$, and   also keeping the lightest mode of fundamental fermions  ${\cal L}^{1}$ for later convenience, we find:  
\begin{eqnarray}
&&{\cal L}^{\rm dual}= {\cal L}^{0} +  {\cal L}^{1} , \qquad   
\cr \cr &&
{\cal L}^{0} = 
 \frac{g^2}{2L} (\partial \sigma)^2 +   \lambda i \bar \sigma_{i}\partial_{i}   \lambda + e^{-S_0} (e^{i  \sigma } \lambda^2 +  {\rm h.c.}) ,
 \label{L0}
   \\ \cr
&& {\cal L}^{1} =  \bar\Psi i (\bar \gamma_{i}D_{i} + i \gamma_4 \langle A_4\rangle)   \Psi  +
 e^{-S_0} \left( e^{- i  \sigma } \lambda^2   \psi_L \psi_R  + {\rm h.c.}\right) +  e^{-2S_0} \left( e^{- 2i  \sigma }    \psi_L \psi_R  + {\rm h.c.}\right)  .\qquad \qquad
 \label{L1}
 \end{eqnarray}  
 
 Interestingly, ${\cal L}^{0}$ is the lagrangian obtained in 3d Yang-Mills theory with adjoint Higgs scalar and adjoint fermion in \cite{Affleck:1982as}. One can examine (\ref{L0}) perturbatively, by expanding around $\sigma$$=$$0$. Since  $e^{i \sigma}$
 is charged under $U(1)_A$, this is equivalent to the spontaneous breaking of  $U(1)_A$. As argued in  \cite{Affleck:1982as}, the photon in this theory remains massless and is, in fact, a  NG boson. The spontaneous breaking of $U(1)_A$  by  $\langle e^{i \sigma}\rangle =1 $  generates a mass term for the adjoint fermion, 
and a 4d-complex mass 
 for the fundamental fermions (which already possess a 3d-chiral symmetric mass term due to $\langle A_4 \rangle$),  given by: 
 \begin{equation}
{\cal L}^{\rm dual} \supset  e^{-S_0}\langle e^{i \sigma}\rangle \lambda^2   + e^{-2S_0} \langle e^{-2i \sigma}\rangle  \psi_L \psi_R   + {\rm h.c.}  =  
 e^{-S_0} \lambda^2   + e^{-2S_0}   \psi_L \psi_R   + {\rm h.c.} 
 \label{fmass}
\end{equation}

{\flushleft{\bf Interpolating between small and large $\mathbf S^1 \times \R^3$:} }
 The picture that emerges when combining the above analysis with the expected behavior on $\R^4$ is that    $U(1)_A$ is broken  at both small and large $L\Lambda$, but  in an unconventional way.  The picture we advocate is shown  below:
  \begin{equation} 
  (N_{\rm adj}, N_{\rm F})=(1,1): \qquad 
 \xy
 (-6,0)*{\bullet}; 
 (-80,0)*{\bullet}; 
(-40,0)*{}; 
(-80,0)**\dir{-} ?(.70)+(4,4)*{   \langle e^{i \sigma}\rangle =1 }; 
(0,0)*{\; \; \;  L}; 
(-40,0)**\dir{-} ?(0.05)*\dir{<} ; 
(-25,3)*{ \langle \lambda \lambda \rangle = \Lambda^3   } ; 
(-5.5,3)*{ {\R^4}} ; 
(-80,3)*{ {\R^3}} ; 
\endxy ~.
\label{phasediag} 
\end{equation}
In other words, the order parameter for $\chi$SB at small $S^1$  is a  topological {\it disorder} operator,  charged under $U(1)_A$  and the  photon is  the corresponding NG-boson. This phenomenon is possible, because the long distance theory is 3d, where photon can be dualized to a scalar. On the other hand, at large $L\Lambda$, we expect $\chi$SB due to a local {\it order} parameter, the  chiral condensates  given in (\ref{chiralc}).  Thus, we conjecture the existence of a Goldstone boson at any  $L$.\footnote{In supersymmetric theories, discrete $\chi$SB by disorder opeartors has already been  observed \cite{Aharony:1997bx} (such as $SU(2)$ SYM on $\R^3 \times S^1$), but we are not aware of supersymmetric examples with continuous $\chi$SB due to disorder operators in the same geometry.}
  
  To see how the small- and large- $\Lambda L$ pictures can merge together, we 
can  extrapolate the dual Lagrangian  (\ref{L0}) and  (\ref{L1}) to  the boundary of its region of validity (i.e., to $\Lambda L \sim 1$) and employ the chiral symmetry breaking induced by the order parameters 
   (\ref{chiralc}). Then we find that  (\ref{L0})  produces:
   \begin{equation}
{\cal L}^{\rm dual} \supset  e^{-S_0} e^{i \sigma}\langle \lambda^2  \rangle + {\rm h.c}
\sim   e^{-S_0}\Lambda^3 \cos {\sigma}~,
\end{equation}
 i.e., a mass term for the dual photon. Thus, the gauge fluctuations in this domain are gapped. At large $L$ the only massless mode is  due to the fluctuations of the chiral condensate parametrized by the pion field.

\subsection{General case of $(N_{\rm adj}, N_{\rm F})$ $SU(N)$ gauge theory}    
 
 Here, we briefly consider theories with $N_{\rm adj} \geq 1$ and $N_{\rm F}$ arbitrary   on  $\R^4$ and  
$\R^3 \times S^1$. The charges of local fields and interesting topological operators can be found by following the
 earlier  analysis. The fundamental fields' charges are: 
  \begin{equation}
  \label{qcdadjf}
\begin{array}{c|ccccc}
& U(1)_B  &  U(1)_{A}  & SU(N_{\rm F})_L &     SU(N_{\rm F})_R  &SU(N_{\rm adj})
 \\ \hline
\lambda & 0 & 1& 1& 1  & \Box \\
\psi_L &  1&  - \frac{NN_{\rm adj} }{N_{\rm F}} &  \Box & 1 & 1 \\
\psi_R &  -1&  - \frac{NN_{\rm adj}}{N_{\rm F}} & 1 &  \Box & 1 \\
\end{array} 
\end{equation}
We will call theories with fermion matter content described above QCD(adj/F).

For confining gauge theories, $N_{\rm F}$-small, at radius $LN\Lambda \ll 1$, these theories exhibit confinement without  $SU(N_{\rm F})_L \times     SU(N_{\rm F})_R  \times SU(N_{\rm adj})$ chiral symmetry breaking. $N-2$ dual photons acquire mass via the magnetic bion mechanism and one remains massless. Similar to the $SU(2)$ example, the massless dual photon has an interpretation as a Goldstone boson of the spontaneously broken $U(1)_A$ and we believe that $U(1)_A$ is broken both at large and small $LN\Lambda$.   At large $NL\Lambda$, the non-abelian part of the chiral symmetry is also expected to be broken down to  
the diagonal $SU(N_{\rm F})_{L+R}  \times SO(N_{\rm adj})$. 
  
 For theories which flow to conformality on $\R^4$,  $U(1)_A$ remains broken at finite $S^1$ and   gets restored in the decompactification limit. 
 
 \subsection{Generalization of  Banks-Zaks CFTs}
 
 The BZ limit corresponds to    a tunably small fixed-point  coupling,  achieved by setting $\frac{N_F}{N} = 5.5(1- \epsilon)$, where $N \rightarrow \infty$ and $\epsilon \ll 1$ (recall that $5.5 N$ is the asymptotic freedom boundary on the number of flavors in QCD(F)). In this domain, the first and second order beta function coefficients balance out and higher orders are suppressed by extra powers of $\epsilon$. Thus, they can be dropped safely.  
 A BZ-limit does not exist for two-index representation fermions, for which the asymptotic freedom boundary   at large-$N$  is $5.5$ and since $N_f$ is integer valued,  the best one can get is a (numerically) weak-coupling fixed point, which is not a parametric smallness. There  is a simple generalization of the BZ-limit in case of mixed representation theories with an admixture of  two- and one-index fermions. Consider the QCD(adj/F) theory (\ref{qcdadjf}).  Then, taking:
 \begin{equation}
 \frac{N_{\rm F}}{N} + N_{\rm Adj} = 5.5(1- \epsilon) ~,
 \end{equation}
one can tune $\epsilon \ll 1$ small for any value of $N_{\rm Adj} \leq 5$ to ensure that the loop expansion parameter at the fixed point can be made parametrically small, $\frac{g_*^2 N}{16 \pi^2} \sim \epsilon$. 

It is also worth noting that among such theories with $N_{\rm Adj}= 5$ and $N_{\rm F}=0.5 N(1- \epsilon)$ with a tunably small   BZ-type fixed point,  an analytic solution of the theory may be given  at any $\R^3 \times S^1$ by employing the twisted partition function.  In this case, the eigenvalues of the Wilson line are approximately uniformly distributed and the semiclassical analysis can me made reliable at any radius. 
 Thus, the concern raised in footnote (\ref{concern}) can be avoided safely as the coupling constant can be made parametrically small. By using the techniques of \cite{Unsal:2007jx, Poppitz:2009uq}, this class of 4d gauge theories can be solved analytically.

 The one-loop potential is:
   \begin{eqnarray}
 \label{oneloopmixed}
   V_{\rm Adj/F}^{+}[\Omega] &=& \frac{2}{\pi^2 L^4} \sum_{n=1}^{\infty} \frac{1}{n^4} \left[ 
   (N_{\rm Adj}-1)  |\tr \Omega^n|^2  +  N_{\rm F} ( \tr \Omega^n + {\rm h.c.})  
   \right]  ~,
\end{eqnarray}
and the corrections are parametrically suppressed due to the tunable  BZ-fixed-point value.  Note that in this potential, the effect of fundamental fermions is suppressed relative to the adjoint fermions and gauge bosons. The contribution of adjoint fluctuations is    $(N_{\rm Adj}-1)  {\cal O}(N^2)$ and the one of 
fundamental fermions is   $N_{\rm F}  {\cal O}(N)$. Since the number of fundamental flavors is also  ${\cal O}(N)$, the second contribution is not suppressed at large-$N$. But despite that, 
since  $\frac{N_{\rm F}/N}  {   (N_{\rm Adj}-1)} \approx \frac{1}{8}$, the back-reaction  of fundamental matter is small. 

 Around the center-symmetric background, we can use semiclassical analysis, which is now reliable at any radius. This gives us a way to  calculate the mass gap for gauge fluctuations:
   \begin{equation}
 \label{massgaplimit2}
m(L)  \leq  \frac{1}{LN} \exp \left[ {- \frac{8 \pi^2} {g^2_*N } } \right] \sim   \frac{1}{LN} \exp \left[ {- \frac{1}{2 \epsilon} }\right]~.
\end{equation}
This is a scale non-perturbatively  suppressed with respect to  $1/LN$. In the decompactification 
limit, the mass gap for gauge fluctuations vanishes, as in a BZ-type CFT at $\R^4$.

  \subsection{Refinement of  the conformality  or confinement criterion}
  \label{refinement}
  
 In previous work, as reviewed in the Introduction, we classified QCD-like theories in four groups according to their mass gap profiles as a function of $L$. From the analysis of \cite{Poppitz:2009uq},  we know that the QCD(F)  theories  with  $N_f < 2.5 N$ are in class-{\bf a}. Since the two-loop beta-function acquires a zero only at  $N_f = 2.61 N$, we argued that the theories in the  range 
  $2.5N<N_f < 2.61 N$ must be  class-{\bf c}. Our next goal is to attempt to determine the  highest value of $N_f$ for  class-{\bf c} theories  and thus to improve our previous estimate of the lower boundary of the conformal window. Of course, by using QCD(F) per se,  this cannot be determined by means of semiclassical techniques.  Instead, in order to make some progress, we will use  a mixed-representation QCD with one adjoint Weyl and $N_f$ fundamental Dirac fermions. 
  
The mass gap for gauge fluctuations in  QCD(adj/F) in the semiclassical domain is generated by magnetic bions. Combining the $N_{\rm adj} = 0$ analysis of  \cite{Poppitz:2009uq} with an analysis similar to the one done for the $SU(2)$ theory above gives:
\begin{eqnarray}
\label{mono}
m_\sigma \sim  
 \frac{1}{L}e^{-\frac{S_0(L)}{2}} = \Lambda(\Lambda L)^{\frac{b_0}{2} -1}, & \qquad  \frac{b_0}{2} -1 = \frac{1}{3}( \frac{5}{2}- 
 \frac{N_{\rm F}}{N} ) \qquad &N_{\rm adj}=0,  
  \\
  \label{bion}
m_\sigma \sim \frac{1}{L}e^{-S_0(L)} = \Lambda(\Lambda L)^{b_0 -1}, &  \qquad b_0 -1= \frac{4}{3}( 4-N_{\rm adj} - 
 \frac{N_{\rm F}}{N} ) \qquad   & N_{\rm adj}\geq 1,
\end{eqnarray}
where $S_0 = 8 \pi^2/(g^2 N)$ is now $1/N$-th of the 4d instanton action.

The reason that (\ref{bion}) does not reduce to (\ref{mono}) is due to the difference of confinement mechanisms: the latter is due to magnetic bions and appears at order $e^{-2S_0}$ and the former is due to monopoles and appears at order $e^{-S_0}$ in the semiclassical expansion.  
It is interesting to note that the critical number of fundamental fermions (determined by the change of mass gap behavior as a function of $L$ at fixed $\Lambda$) {\it increases} to $N_{\rm F}$$=$$3N$ once  $N_{\rm adj}$$=$$1$. Since adding more fermions enhances the screening effects, it is not expected to  increase the critical number of flavors and we are led to conclude that some QCD(F) theories with $N_f > 2.61 N_c$ are also class-{\bf c}. 
If we now use 1-adjoint as $N$ fundamentals for perturbative purposes, it is then possible that all QCD(F) theories in the   $2.5N$$<$$N_{\rm F} < 4N$ range  are class-{\bf c}.  
Now, it is of course possible that  QCD(adj/F) theories with $(1, 2.5N)$$<$$(N_{\rm adj}, N_{\rm F})$$<$$(1,3N)$ themselves exhibit 
   class-{\bf d} behavior, in particular if $N_{\rm F}$ is close to $3N$. Thus, their ``image" QCD(F) theories are expected to be class-{\bf b}
   (since QCD(F) for $N_F$ near $4N$ has, at small $\Lambda L$, a mass gap decreasing with $L$). 
    Thus, the best we can argue is  that  
$N_{\rm F} = 4N$ is an upper bound for the lower boundary of conformal window. 
We thus argue that, for QCD(F), class-{\bf c} theories are in the domain  $2.5N$$<$$N_F$$<$$4N$  and the class-{\bf c} domain ends up before it hits  the $4N$ limit.

Eqn.(\ref{bion}) also provides estimates for gauge theories with mixed-representation fermions, as in the  conformal house of \cite{Ryttov:2009yw}. We note that the result  from deformation theory is much closer to the $\gamma=1$ estimates of   ref.~\cite{Ryttov:2009yw} than to the  $\gamma=2$ ones.

\begin{table}[ht]
\scriptsize
\begin{center}
\begin{tabular}{|c |c |c|c|c|c| }
\hline
  $N$  & D.T. 1a/1(a+c)\qquad & Ladder (SD)-approx.& Functional RG & NSVZ-inspired: $\gamma = 2$/$\gamma = 1$  & $N_{\rm F}^{AF}$ \\ \hline
  2 & $ 5/8 $ &    $7.85 $  & $8.25$  & $5.5$/7.33& 11        \\ \hline
  3 & $7.5/12 $ &   $11.91  $  &  $10$& $8.25$/11 & 16.5   \\ \hline
    4 &  $10/16 $          & $15.93$ &  $13.5$ &$11$/14.66 & 22           \\ \hline
   5 &   $12.5/20 $  &   $19.95  $   &$16.25$  &$13.75$/18.33 & 27.5  \\ \hline
    10 &   $25/40 $    &  $39.97 $  & n/a &$ 27.5$/36.66 & 55  \\ \hline
        $\infty$ & $ 2.5 N$/$4N$  &  $  4 N$& $\sim (2.75-3.25)N$&$ 2.75N$/$3.66N$   & $5.5N$ \\ \hline
\end{tabular}
\end{center}
\caption{Estimates  for the lower boundary of conformal window for  QCD(F), $N_{\rm F}^* < N_{\rm F}  < 5.5 N$. The results of the deformation theory approach according to \cite{Poppitz:2009uq} are shown in the ``D.T. 1a" column, while those due to the ``refined" estimate of this paper are shown under ``D.T. 1(a+c)". } 
\label{defaultF}
\end{table}%
 \begin{figure}[h] 
\centering
\includegraphics[width=5in]{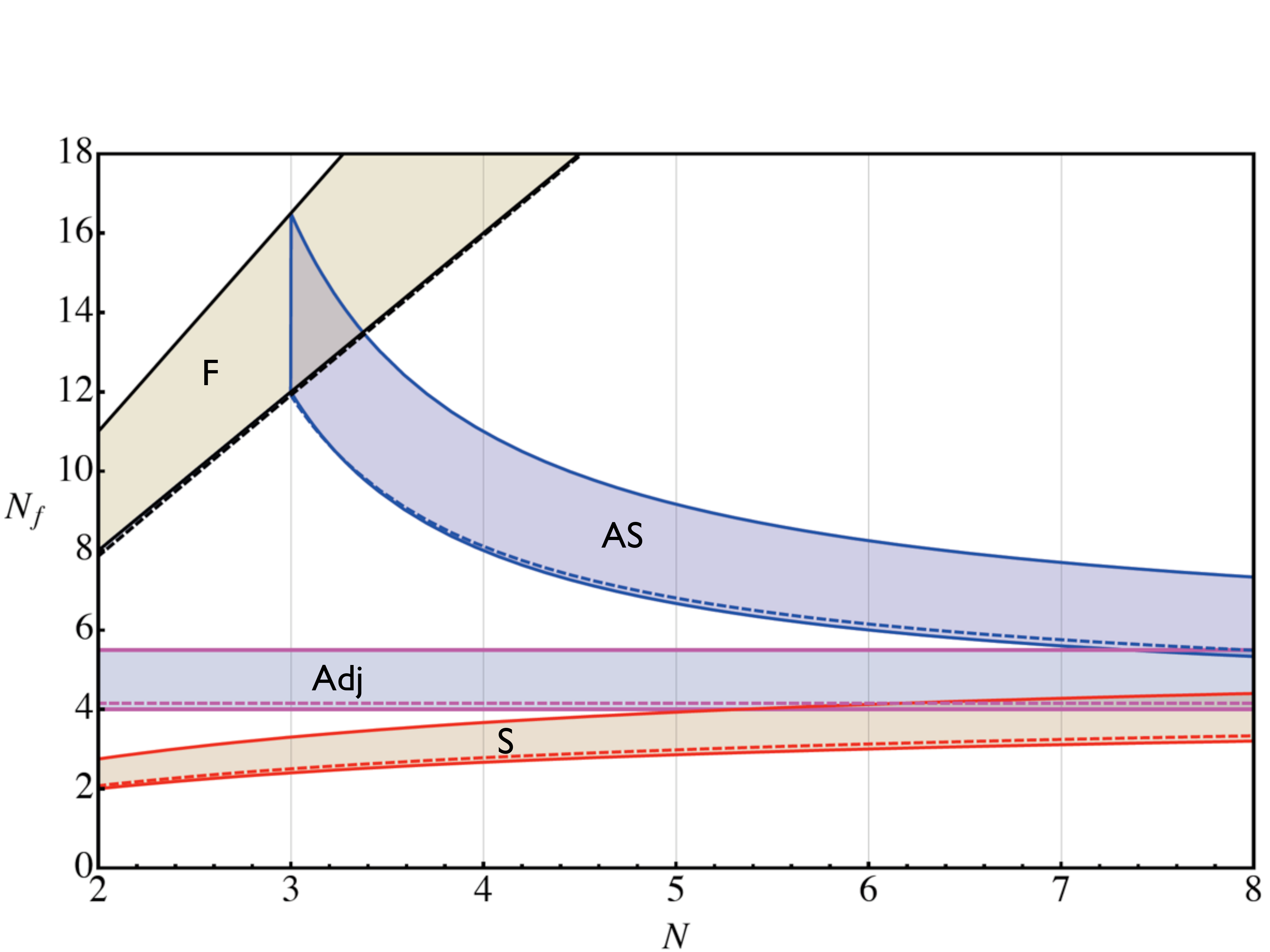}
\caption{Conformal window estimates for QCD(F/AS/Adj/S) by using  deformation theory and the mass gap criterion of this paper 
(solid lines, upper limits on the lower boundary) and the truncated Schwinger-Dyson approximation (dashed-lines). In order to not overcrowd the figure, we do not plot the estimates of other approaches; see Table~1.}
\label{improvefigure}
\end{figure}

In Table~\ref{defaultF}, we tabulate the estimate of the lower boundary of the conformal window  theories according to this ``refinement" of our previous criterion, which we also show in the table.
The reader should bear in mind that the estimate $N_{\rm F} = 4N$ is only an upper bound on the lower boundary of conformal window. 
For convenience of the reader, we have also listed the estimates of the ladder approximation to the Schwinger-Dyson equations 
\cite{Appelquist:1996dq,Miransky:1996pd,Appelquist:1998rb,Dietrich:2006cm},   NSVZ-inspired beta-function conjecture \cite{Ryttov:2007cx} (see also \cite{Antipin:2009wr}), and functional renormalization group approach \cite{Gies:2005as}. Estimates were also obtained by  using properties of the multi-loop beta function \cite{Gardi:1998ch, Chishtie:1999tx},  via a conjectured thermal inequality \cite{Appelquist:1999hr} (see also discussion in \cite{Sannino:2005sk}),
via the worldline formalism \cite{Armoni:2009jn}, and from a conjectured dual of QCD(F) \cite{Sannino:2009qc}. It is also of some interest to show the estimates of the ladder approximation and 
our approach based on mass gap for gauge fluctuations for theories with $N_f$ flavors of Dirac fermions in the fundamental (F) and two-index representations---antisymmetric (AS), symmetric (S), and adjoint (Adj).  The results are plotted\footnote{The Schwinger-Dyson estimates for two-index representations are taken from  \cite{Dietrich:2006cm, Ryttov:2007cx}. We thank T. Ryttov and F. Sannino for sharing their Mathematica file with us.}  on Fig.~\ref{improvefigure}.  We also note that the recently conjectured higher-representation dual of \cite{Sannino:2009me}, like our ``refined" estimate, also gives results agreeing with the Schwinger-Dyson equations.  For studies of conformal windows in $SO$ an $SP$ gauge groups, see \cite{Sannino:2009aw, Golkar:2009aq}.

The results plotted on the figure should look surprising: the  ladder approximation does not know anything about confinement and uses the two-loop beta function and the one-loop result for the fermion-bilinear anomalous dimension. Our proposal uses the $L$-dependence  of  mass gap for gauge fluctuations as an identifier, and employs semiclassical techniques along with the one-loop beta function for dimensional transmutation.  It is  quite surprising that these two ways to estimate conformal window produce such close estimates.

\section{Topological disorder operators and  chiral symmetry breaking }
\label{disorderchisb}

Chiral symmetry in ordinary QCD with fermions in a single (non-mixed) representation are of two types: non-abelian continuous chiral symmetry  and abelian discrete chiral symmetry. 
The classical $U(1)_A$ axial symmetry is reduced to $ \Z_{2hn_f}$  due to instantons, where 
$2hn_f$ is the number of zero modes in the instanton background. 
Theories with mixed representation fermions have two types of classical axial symmetry 
$U(1)_{A_1} \times U(1)_{A_2}$, reduced by instanton effects to a single
$U(1)_A$,  similar to SUSY-QCD.  Thus, the chiral symmetries can generally be written as: 
\begin{equation}
G_\chi= G_{\rm non-ab.} \times   G_{\rm ab.}  = \left\{ \begin{array}{ll}  
G_{\rm non-ab.}  \times \Z_{2hn_f} & \qquad  {\rm  pure \; rep.}  \\
G_{\rm non-ab.}  \times U(1)_A & \qquad  {\rm  mixed  \; rep.}  .
\end{array} \right.
\end{equation}
In what follows, this distinction between non-abelian and abelian chiral symmetry is particularly useful.

In the analysis of gauge theories on $\R^3 \times S^1$,  one of the interesting phenomena that we have learned  is that the pure flux operators in QCD-like theories are  charged under the   $G_{\rm ab.}$, but neutral under   $G_{\rm non-ab.}$.  
This means that, apart from chiral fermion condensates,  the pure monopole operators  such as $e^{i \sigma}$---the  topological disorder operators which cannot be locally expressed in terms of fields of the microscopic theory---are also good order parameters for the chiral symmetry. 
 In fact, in  \S\ref{solvable},  we  showed that the $\Z_{20}$ chiral symmetry of the $SU(2)$ gauge theory with one fermion in 4-index representation   is broken down to $\Z_2$ by the condensation of a pure flux operator, $\langle e^{i \sigma} \rangle= e^{i \frac{2 \pi k}{N}}, k=1, \ldots, 10$.  In mixed representation theories, $\langle e^{i \sigma} \rangle= e^{i \xi}, \
 \xi \in[0, 2 \pi) $.  
 
 The abelian $\chi$SB,  
 either discrete or continuous, induces  mass  for fermions in theories without non-abelian chiral 
 symmetries, i.e, the theories for which $G_{\rm non-ab.}$ is trivial. These are one-flavor QCD-like theories  or mixed representation theories with $(N_{\rm adj}, N_{\rm F})=(1,1)$.   In theories for which  $G_{\rm non-ab.}$ is non-trivial, the spontaneous breaking of    $G_{\rm ab.}$ does not induce a mass gap for fermions.   
 
 Let us explain, in generality, how the  $ G_{\rm ab.}$  breaking takes place in the small 
 $S^1 \times \R^3 $ regime.  $ G_{\rm ab.}$ is an exact symmetry of the microscopic theory, 
 thus it must be an exact symmetry of the long-distance effective theory (otherwise, it would be    anomalous, which is incorrect).  Consider a typical monopole operator with its 
 fermion zero-mode insertions.  The zero-modes structure is manifestly invariant 
 under  $G_{\rm non-ab.}$, but rotates under   $ G_{\rm ab.}$. 
 In the long-distance theory  without monopoles, there is also an infrared 
  topological $U(1)_J$     symmetry, which shifts the dual photon. The $U(1)_J$ symmetry intertwines with the $G_{\rm ab.}$ to render the monopole operator invariant:
 \begin{equation}
  [G_{\rm ab.}  \times U(1)_J] \rightarrow  [G_{\rm ab.}]_{*} ~.
 \end{equation}
 The abelian shift symmetry $[G_{\rm ab.}]_{*}$  forbids all flux operators which are not invariant, and since it is intertwined with the topological $U(1)_J$, it may be referred to as  {\it topological shift  symmetry}.    In fact, all  the bosonic potentials induced, typically, by non-selfdual topological excitations  
 in the semiclassical regime,  obey:
    \begin{equation}
 [G_{\rm ab.}]_{*} :    V^{\rm np} (\mbox{$\sigma$}) \rightarrow  V^{\rm np} ([G_{\rm ab.}]_{*}  \mbox{$\sigma$})  = V^{\rm np} (\mbox{$\sigma$}) ~,
 \end{equation}
 for both discrete and continuous  abelian chiral symmetries.  For the discrete symmetry case, the topological shift symmetry $[G_{\rm ab.}]_{*}$ connects the $h$ isolated vacua, while  for continuous $U(1)_A$ it implies an $S^1$ vacuum manifold. 

A few examples would be helpful. Consider one-flavor global anomaly-free $SU(2)$ theories with $j= 1, \frac{3}{2}, 2$ fermions.   Then, the bosonic potentials are, respectively:
\begin{equation}
\cos (2 \sigma), \;  \cos (5 \sigma), \;  \cos (10 \sigma)~,
\end{equation}
leading to two, five, and ten  isolated vacua on $\R^3 \times S^1$ (as we have seen in 
\S\ref{solvable}, the last case has runaway vacua, which means that they move to infinity in the decompactification limit).  For continuous $\chi$SB,  a good example is  discussed 
 \S\ref{mixed1}, where  $V^{\rm np} (\sigma) = {\rm constant}, \sigma \in [0, 2 \pi)$.   The  vacuum expectation values and breaking patterns are, thus: 
 \begin{equation}
\left\{ \begin{array}{llll}    \langle e^{i \alpha_i \sigma} \rangle= e^{i \frac{2 \pi k}{N}}, \; 
& k=1, \ldots, h &  \qquad \Z_{h}  \rightarrow \Z_1
  & \qquad  {\rm  pure \; rep.}  \\
\langle e^{i  \sigma} \rangle= e^{i \xi},  \; 
& \xi \in[0, 2 \pi)  &  \qquad   U(1)_A \rightarrow  \Z_1  & \qquad  {\rm  mixed  \; rep.}  
\end{array} \right.
\label{condensate}
\end{equation} 
We reach to the following conclusions regarding abelian chiral symmetry breaking  on small $S^1 \times \R^3$:
\begin{itemize}
\item The dynamical breaking of the abelian chiral symmetry, either continuous or discrete, upon a duality transform, maps into a spontaneous breaking by a tree level potential. The potential is  induced by non-selfdual topological excitations. 
\item  This phenomenon occurs in a weakly-coupled domain where the anomalous dimension of fermion bilinears is small $ \gamma(L) \ll 1$ and the  truncated Schwinger-Dyson (ladder) approximation would {\it not} induce  $\chi$SB. 
\item In theories with  trivial $G_{\rm non-ab.}$, the condensation $\langle e^{i  \sigma} \rangle\neq 0$ of  topological excitations with unit magnetic charge is capable of generating 
a 4d-complex mass for fermions. 
\end{itemize} 
It is usually accepted  that dynamical symmetry breaking (DSB) is a difficult nonperturbative phenomenon and that spontaneous breaking (SSB) by a potential is relatively simple. Remarkably,  the duality transformation maps the problem of DSB to SSB by a tree level potential.\footnote{It is commonly believed that if 
YM theory or QCD-like theories could ever be solved by duality, 
nonperturbative phenomena such as mass  gap for gauge fluctuations and chiral symmetry breaking would be tree-level effects in the dual formulation. Indeed, this dream finds realization in one-flavor QCD-like theories in the calculable small $S^1 \times \R^3$ domain. }

The second result is still surprising, but is somehow expected in the light of the first one. SSB by a potential is a phenomenon that can happen at weak coupling as well. 
 $\chi$SB is induced in the ladder approximation if 
$\gamma _{\bar \psi \psi} \simeq 1$ is reached. By squeezing the space  to the perturbative domain 
$L \ll \Lambda^{-1}$ and due to gauge symmetry breaking by the nontrivial holonomy,  the theory is engineered to 
 remain in the  $ \gamma _{\bar \psi \psi}(L) \ll 1$ domain for small $S^1$. Thus, chiral symmetry cannot break to all orders in perturbation theory within this domain.   The solution to the gap equation would yield zero and the theory does not generate a mass term for fermions.   These are shortcomings of perturbation theory, as the topological disorder operators, which are part of the  dynamics and whose condensation can generate mass gap for fermions, do not appear perturbatively. 
 
 In the small $S^1$ domain, the condensate $\langle e^{i  \sigma} \rangle\neq 0$ also has an interesting physical interpretation.  $e^{i  \sigma(x)}$ corresponds, in the original electric theory, to the insertion  of magnetic charge at point $x\in \R^3$. Thus, in the dual formulation of the theory, $\langle e^{i  \sigma} \rangle\neq 0$  is a vacuum condensate of magnetically charged excitations. As shown in (\ref{fmass}), such a condensate is capable of giving a 4d-complex mass 
to the fermions in all non-abelian gauge theories for which  $G_{\rm non-ab.}$ is trivial.  Needless to say, since $ \gamma _{\bar \psi \psi}(L) \ll 1$ in this regime,  $G_{\rm non-ab.}$ is not broken either.  In fact, in center-symmetric gauge theories, $LN\Lambda \ll 1$ is the domain of confinement without non-abelian continuous $\chi$SB.

 \acknowledgments We  thank  D.T. Son  for useful discussions.  
 This work was supported by the U.S.\ Department of Energy Grant DE-AC02-76SF00515 and by the National Science and Engineering Council of Canada (NSERC).

\section{Addendum: Mixed chiral/vectorlike representations}
After this paper was submitted for publication, Ref.~\cite{Sannino:2009za} appeared, studying  conformal windows of theories with mixed chiral and vectorlike representations  
  in the framework of the proposed all-orders beta function
(for older work on this subject based on the thermal inequality, see Refs.~\cite{Appelquist:1999vs, Appelquist:2000qg}).
Since mixed-representation theories are  considered in this paper, 
in order to facilitate comparison between the results of different approaches, we now give the estimates of   deformation theory for the conformal windows in these theories. We feel that a comparison is important to the better understanding of the various approaches and in their development.\footnote{We do not give many details, as the analysis of the dynamics of the chiral and vector-like theories on $\R^3 \times S^1$ is very similar to the analysis of other models in this paper and in \cite{Poppitz:2009uq}. }

We consider three classes of mixed chiral/vector-like representation theories.  These theories 
are composed of  $N_g$ chiral ``generations"  and  $N_f$ vectorlike generations of massless fundamental Dirac fermions. These three classes are:\footnote{The Type(A) $N_g=1$ theories with arbitrary $N_f$ is also called a ``generalized-GG"-model, whereas the  Type(S) $N_g=1$ theories with arbitrary $N_f$ is sometimes called 
a ``generalized-BY"-model  \cite{Sannino:2009za}.}
\begin{itemize}
\item Type(A) $\qquad   N_g[ AS,  (N-4) {\overline F} ]  + N_f(F, {\overline F})$
\item Type(S)  $\qquad   N_g[ S,  (N+4) {\overline F} ]  + N_f(F, {\overline F})$
\item Chiral    quivers: $\bigoplus_{J=1}^K  \left[ N_g 
(1, \ldots, N_J , \overline{N}_{J+1}, \ldots 1) +   N_f (1, \ldots, (F, {\overline F})_J, \ldots 1)   \right] $
\end{itemize}
The gauge group for the first two classes is $SU(N)$ and for the chiral quivers, it is  $SU(N)^K$.  

In a mixed chiral/vector-like gauge theory of the above type and  for a generic value of $N$, the mass gap for gauge fluctuations is induced  predominantly by magnetic bions. This data can be extracted by using  the structure of the fermionic  zero modes of monopole operators and can be found in \cite{Poppitz:2009uq}.  According to our criterion, as discussed in the Introduction,  if the mass gap for gauge fluctuation is an increasing function of $L$ for a given $(N_g, N_f)$, we claim that such theories exhibit confinement in the $\R^4$ limit;  if the mass gap in gauge sector is a decreasing function which asymptotes to zero, then we expect such theories to flow to conformality in the $\R^4$ limit. (Certainly, by the definition of interacting CFTs on $\R^4$, massive gluons are not acceptable.) 
 
 Taking the bion contribution as the leading one, we find that the mass gap for gauge fluctuations is:
 \begin{equation}
 \label{GGbions}
 m_\sigma \sim {1\over L} e^{- S_0} = \Lambda (\Lambda L)^{8 N - 2 N_f - 2 N_g (N- 3a) \over 3 N}~,
 \end{equation}
 where $a=1$ for Type(A), $a=-1$ for Type(S), $a=0$ for chiral quivers. The mass gap  decreases as $L$ increases for $N_f + (N-3a)N_g > 4 N$.   Thus our estimate for the lower boundary of the conformal window for these  theories are:
  \begin{eqnarray}
&&{\rm Type(A)}: \qquad  N_f^* + (N-3) N_g^* = 4 N~.  \cr
&&{\rm Type(S)}:  \qquad   N_f^* + (N+3) N_g^* = 4 N~. \cr
&&{\rm Chiral \; quivers}:   \qquad   N_f^* + N N_g^* = 4 N~.
\end{eqnarray}
Few comments are in order: For $N_g=0$, this result gives us our estimates for the lower boundary in QCD(F) theory. For   $N_f=0$, the purely chiral case, these estimates are the one 
given in our earlier work  \cite{Poppitz:2009uq}. 

The Type(A) and Type(S) theories with 
$N_g=0, 1$ 
are recently analyzed in the framework of the proposed all-orders-beta function  in \cite{Sannino:2009za}.  That analysis can be generalized easily to all the theories that we discuss above. If one uses  $\gamma =1$ (for vector-like matter)  in order to determine the conformal window boundary, one finds 
\begin{eqnarray}
3N_f^* + 2 N_g^*(N-3a) = 11 N
\end{eqnarray}
 where $a=\pm,0$ for the three class of theories.  For the purely vector-like case, this gives 
the  $\gamma=1$ estimate  of \cite{Ryttov:2007cx}, while for $N_g=1$, it gives  
$N_f^* = 3N \pm 2a$, which is very close to our  $N_f^* = 3N \pm 3a$.

 However, despite  this close  agreement between  our result in the vector-like or mostly vector-like cases, there is a sharp disagreement of results as one approaches to purely chiral $N_f=0$ (or mostly chiral) theories. Our formalism,  in the pure chiral case, predicts the existence of a conformal window, starting at 
$N_g^*= \frac{4N}{N-3a}$, whereas the asymptotic freedom boundary is located at 
$N_g^{AF}= \frac{11N}{2(N-3a)}$.  The formalism of  \cite{Sannino:2009za} predicts that $N_g^*= \frac{11N}{2(N-3a)}$, in other words, the absence of conformal window in purely chiral theories. 

The reason that  Ref.~\cite{Sannino:2009za}  predicts the absence of a conformal window for purely chiral theories can be seen in the proposed all-orders beta function, where the anomalous dimension of a fermion bilinear enters. However, in a purely chiral theory, there are no such gauge invariant fermion bilinear.  This leads  Ref.~\cite{Sannino:2009za}  to the conclusion that 
conformal window for such theories should coincide with the vanishing of the first coefficient of the beta function. 

 We think that it is natural to expect that purely chiral ``multi-generation" theories can have conformal windows. To this end, we note  the existence of a  large-$N$ orbifold equivalence between multi-adjoint vectorlike theories and   their $\Z_K$ orbifold projections, which are 
  chiral quiver theories, i.e., $SU(NK)  \longrightarrow  SU(N)^K $.
   The non-perturbative validity of this  equivalence relies on unbroken 
 discrete  $\Z_K$   chiral symmetry in the parent QCD(adj)  and unbroken discrete  $\Z_K$  translation  symmetry of the daughter  chiral quiver theory.\footnote{This condition is obviously violated for the QCD(adj) theories in the confining domain. Thus, there exist no large-$N$ equivalence  between confining QCD(adj) and its chiral daughters \cite{Kovtun:2005kh}. Whereas, above, we suggest that there may indeed be such equivalences between conformal  vector-like and chiral theories.}    It is currently believed that QCD(adj) theories indeed possess a conformal window. If true, this implies the absence of discrete chiral symmetry breaking for the QCD(adj) theories in the conformal window. Thus, the validity of the equivalence relies on the $\Z_K$  translation  symmetry of the daughter quiver theory. Currently, there exist no evidence which may suggest this latter symmetry is spontaneously broken. Thus, it is plausible, but unproven, that there exist a non-perturbative equivalence between conformal QCD(adj) and conformal chiral quiver theories.  
If true, this   requires that multi-generation chiral quivers a conformal window should exist (as found with the present approach in \cite{Poppitz:2009uq}), as it does in the multi-adjoint theories.
For chiral theories of Type(A) and Type(S), there exist no such useful large-N equivalences which relates them to vector-like theories. However, currently there is no evidence that may suggest the absence of a conformal window, apart from the  proposed all-order beta function.   

\bigskip
 
\appendix

\section{Mass gap of a theory versus mass gap for its gauge fluctuations}
\label{massgap}
\label{A}

In ref.\cite{Poppitz:2009uq}, the presence versus absence of a mass gap for the gauge fluctuations of a gauge theory is proposed as an invariant characterization of confinement versus 
conformality in non-abelian gauge theories with (chiral or vectorlike) fermionic matter. 
It is also stated that the calculation of this quantity is out of reach on $\R^4$ with the present understanding of chiral or vectorlike gauge theories. 
The recent  progress allows its  computation on $\R^3 \times S^1$.

In this appendix, we will carefully distinguish  the notion of presence vs. absence of mass gap for gauge fluctuations  from the presence versus absence of massless states  in the Hilbert space of the gauge theory.

Appendix  \ref{A} is a summary of mostly known aspects of gauge theories  and follows very closely  Lecture 16 of Witten's ``Dynamical Aspects of QFT" lectures in \cite{Deligne:1999qp}.  Our addition to this classification is very small---the existence of an exceptional case in one-flavor theories which flow  to CFTs and stating the existence  of confining theories with both NG-bosons and massless fermions. The last class was conjectured as a possibility by 't Hooft, and up to our knowledge, the examples we provide are the first examples of this type in non-supersymmetric gauge theories.

\subsection{Theories without continuous global symmetries}
{\flushleft{${\bf N_f=0}$:}}  Let  two Wilson loops $C_1$ and $C_2$ be  embedded into an $\R^3$ submanifold of  $\R^3 \times S^1$ and the separation between the two be labeled as $ {d}(C_1, C_2)$ (we assume  $ d(C_1, C_2)$ is much larger than the typical size of $C_i$'s).  In the deformed-YM theory  (YM$^*$) on  $\R^3 \times S^1$, their connected correlator is: 
\begin{eqnarray}
\langle W(C_1) W(C_2) \rangle(L)  = e^{-m_g(L) d(C_1, C_2)} + \ldots \qquad  {\rm YM}^*~,
\end{eqnarray}
where $m_g(L)$ is the mass gap profile of the theory as a function of $L$.  The ellipsis  stands for excitations with higher mass that can be exchanged between the two loops, which are suppressed exponentially. 
 $m_g(L)$ is expected to saturate to its $\R^4$  value  for $LN\Lambda \gg 1$ even if $L \Lambda \ll 1$ as a 
result of the  volume independence theorem  of YM$^{*}$ theory \cite{Unsal:2008ch}:
\begin{equation}
m_g|_{{\rm YM}, \; \R^4}= m_g|_{{\rm YM}^{*}, \; LN\Lambda \gg 1} ~.
\end{equation}

{\flushleft{${\bf N_f=1}$:}} In QCD$({\cal R})^*$ theories without continuous global  symmetries,  the form of the connected correlator  is again the same.  The leading term will be due to exchange of the 
massive excitations, and perhaps  an ``eta-prime."  

{\flushleft{\bf Exception:}} Theories with $N_f \leq 1$ are usually 
believed to  possess a mass gap. An exception is the class of multi-index   one-flavor CFTs discussed in this work.

\subsection{Confining theories with continuous global symmetries}

{\flushleft{${\bf 2 \leq  N_f < N_f^*}$}:} There are remarkably powerful theorems  for this class of theories, mainly put forward by 't Hooft.
Let the anomaly-free global chiral symmetry of the theory be $G_\chi$, and its associated currents be $J^a, \;  a=1, \ldots, {\rm dim}(G_\chi)$.  The idea is to use the short-distance physics to constrain the massless  degrees of freedom of the long-distance physics  by considering the anomalies these symmetries would have if they were gauged. Since anomalies are an {\it all-scale} property of a gauge theory, it is possible to  prove the presence of gapless spin-$0$ or spin-$\half$ excitations in the IR theory. 

The ``'t Hooft anomalies"  of the short-distance theory can be extracted from the 3-point functions of the currents  $J^a$ and are encoded in  the invariant cubic tensor 
$d^{abc}= \half \tr T^a\{ T^b, T^c\}$ 
where $T^a,  \;  a=1, \ldots, {\rm dim}(G_\chi)$, are the generators of $G_\chi$, and the trace is taken over the massless (in the UV) fermion representations.  There are three cases allowing the short distance anomalies to be reproduced by  massless IR-degrees of freedom:
\begin{itemize}
\item[{\bf i)}] only  by massless NG-bosons (by the tree-level IR chiral Lagrangian),
\item[{\bf ii)}] only by massless composite fermions (by an IR-loop effect), 
\item[{\bf iii)}] by a combination of massless spin-0 and spin-$\half$ particles (tree-level for spin-0 and loop effect for spin--$\half$).
\end{itemize}
 We believe that all three cases find realization in   non-supersymmetric vectorlike or chiral gauge theories.  More explanation and examples of each class are given below. 

We assume that the theory  has a generic chiral symmetry breaking pattern:
\begin{equation}
G_\chi \longrightarrow \widetilde G\,~. 
\end{equation}
\begin{itemize}
\item[{\bf i)}]  If $\widetilde G$ is the maximal vectorlike subgroup of $G_\chi$,  
anomalies restricted to  $\widetilde G$ vanish, $d^{abc}|_{\widetilde G}=0$. In the infrared, $J_{5, \mu}^a \sim \partial_{\mu} \pi^a$ and the $d$-tensor of the UV theory can be reproduced by pions $\pi^a$ of the broken chiral symmetry $G_\chi/\widetilde G$, similar to the chiral limit of QCD. The masslessness   of the NG-bosons is protected by broken symmetry and the Goldstone theorem.  
\item[{\bf ii)}] If $\widetilde G= G_{\chi} $, the theory exhibits    confinement without chiral symmetry breaking and 
 there are  no Goldstone bosons.  The UV anomalies are reproduced by massless composite fermions. The masslessness of the fermions is due to the unbroken chiral symmetry. 
Examples are the $SU(N)$ chiral theory with  one anti-symmetric (symmetric) and $N$$-$$4$$(N$$+$$4)$ antifundamental left handed-Weyl fermions. 
\item[{\bf iii)}]  It is possible that $\widetilde G$ be a proper subgroup of $G_\chi$ and 
 $d^{abc}|_{\widetilde G}\neq 0$, i.e, $\widetilde G$ is still chiral.  
   The first non-supersymmetric examples (we are aware of) in this class are 
provided by the multi-generation $2 \leq N_W \leq N_W^*$ generalization of the chiral gauge theories given above.\footnote{These theories are discussed in \cite{Poppitz:2009uq}. The anomaly matching is not studied in literature (however, see the most attractive channel study of related theories in \cite{Terning:1994sc}). Nonetheless, we checked that  if one assumes $\widetilde G= G_{\chi}$, 
 the anomalies cannot be saturated by the massless fermions. This implies that $\widetilde G$
 must be a  proper subgroup of $G_{\chi}$ and anomalies must be saturated by  both 
 Goldstone bosons and  fermions.   }
\end{itemize}
The  IR limits of these three classes of confining gauge theories are free. The long distance physics is described by massless bosons, massless fermions, or both. Thus,  the connected correlator of the two-well separated Wilson loops will be dominated by the massless long-distance modes:
\begin{eqnarray}
\langle W(C_1) W(C_2) \rangle|_{\R^4} = ({\rm free \; IR}-{\rm limit \; contribution}) + e^{-m_g d(C_1, C_2)} + \ldots \qquad .
\label{concor}
\end{eqnarray}
In the expansion, the exchange due to glue (gauge fluctuations)  will be extremely suppressed, but is nonetheless present in the full theory.  The exponent of this sub-leading term is what we mean by ``mass gap for gauge fluctuations" in this class of theories. We should, however, admit that this notion is not unambiguously defined on $\R^4$; for example, in QCD $m_g$ receives contributions of pairs of $\rho$-mesons as well as glueballs. However, as we explain in the following Appendix B on $\R^3 \times S^1$, this notion can be made unambiguous at least for small $L$.

\subsection{Conformal  theories with continuous global symmetries}

${\bf  N_f^* < N_f< N_f^{AF}:}$ These are asymptotically free vectorlike or chiral theories, 
whose IR limit is an interacting CFT.  In this class, the 't Hooft anomaly matching is trivially satisfied. This class of theories break no global symmetries. In these theories, both fermions as well as  gauge fluctuations remain massless non-perturbatively. Thus, in the 
connected correlator (\ref{concor}) for such theories, there are no dynamically generated scales, i.e, $m_g=0$.

\section{Wilson loop correlators on $\R^3 \times S^1$ and mass gap for gauge fluctuations}

It is well-known that Wilson (order) and 't Hooft (disorder) operators (or  their more elaborate  refinements)  may be used in determining the infrared behavior, and may in principle be used to determine conformal  vs.  confining behavior of an asymptotically free gauge theory. Although it is implicitly known that the mass gap for gauge fluctuations may also serve the same goal,   the problem is almost never stated in these terms. The reason is clear: It is hard to isolate this observable from a correlator. A typical gauge theory with 
$2 \leq N_f \leq N_f^*$ has an IR-free description in terms of NG-bosons and/or fermions and  such IR descriptions ``forget" about the massive glueball-like particles.  This, however, does not mean that these states are absent in the full theory;  we would like to take advantage of that. 

In the recently developed twisted partition function and/or deformation theory formalism,  it is  understood that  the mass gap for gauge fluctuation is  indeed {\it calculable} on $\R^3 \times S^1$ as a function of $L$  in a domain where semiclassical analysis is reliable. This is true irrespective of the chiral or vectorlike nature of the theory. The limit $L \rightarrow \infty$ of 
$m_\sigma(L)$ determines conformal or confining behavior. 

Now, we present some ideas on how to extract the profiles shown in Fig.~\ref{fig:massgap}, for example, using lattice gauge theory.\footnote{In what follows,  QCD(adj) is not included as confinement  already has a precise  definition in terms of Wilson loops in the defining representation. Our goal below is to provide such an  unambiguous definition for all QCD-like or chiral theories with complex representation fermions, such as fundamental. The mass gap in the gauge sector serves this goal.}
  It is necessary to use an asymmetric lattice 
$T^3 \times S^1$, which mimics $\R^3 \times S^1$.  With this in mind, the discussion can also be given in the continuum. 
We stabilize the center symmetry to  guarantee smoothness in the sense of (approximate) center symmetry. Then,  assume the theory under consideration has some global  
anomaly-free $U(1)_{B \; {\rm or } \; A}$ (vectorlike or chiral) symmetry.  In the weakly coupled domain of any of such theories, 
the Wilson line behaves as an adjoint Higgs field, and typically provides a real, chiral-symmetric 3d mass term for the fermions. For a generic Weyl fermion, this is:
\begin{equation}
\overline \psi \overline  \sigma^{\mu} D_{\mu} \psi \rightarrow 
\overline \psi ( \overline  \sigma^i D_i + i  \overline  \sigma^4 \langle A_4 \rangle) \psi
\label{Dirac1}~.
\end{equation}
Most fermionic matter will acquire mass of order $1/L$ due to the    $\langle A_4 \rangle$ vev. 
If there are massless modes left over, we use boundary conditions with either a $U(1)_B$ or 
$U(1)_A$ twist:
\begin{equation}
\psi(L) = e^{i \alpha} \psi(0)~.
\label{twist}
\end{equation}
By a field redefinition $\psi' (x, x_4)= e^{-i \frac{\alpha x_4}{L}} \psi(x,x_4)$, the  
periodicity of the fermion may be regained,  while $\frac{\alpha}{L} \equiv m$ is absorbed 
 into the Dirac  operator as a 3d  chiral-symmetric real mass term. Thus, (\ref{Dirac1}) becomes: 
\begin{equation}
\overline \psi' [ \overline  \sigma^i D_i + i  \overline  \sigma^4 ( \langle A_4 \rangle + m)] \psi'
\label{Dirac2}~.
\end{equation}
The point of this manipulation\footnote{We stress that the twist in (\ref{twist}) has to correspond to an anomaly-free $U(1)_{A \; {\rm or} \; B}$ symmetry, to avoid the generation of  Chern-Simons terms which alter the infrared dynamics \cite{Poppitz:2008hr}. }  is to give chirally-symmetric masses of order $1/L$ to all the the degrees of freedom but the   gauge bosons in the Cartan subalgebra; note, however, that additional chiral violating fermion mass terms can be generated by topological disorder operators as described in Section \ref{disorderchisb}.

The twist (\ref{twist}) does not alter the mechanisms of confinement for gauge theories on $\R^3 \times S^1$, in particular the analysis of refs.~\cite{Shifman:2008ja,Poppitz:2009uq} remains valid. Since all center-stabilized gauge theories with $N_f \leq N_f^{AF}$ fermions have a semiclassical domain 
with confinement without chiral symmetry breaking and since, in this domain, we can lift 
fermion zero modes by using judiciously chosen $U(1)$ twists, the only light modes are the ones associated with the gauge sector, and the following relation holds:
\begin{eqnarray}
\langle W(C_1) W(C_2) \rangle|_{\rm twist} (L)  = e^{-m_g(L) d(C_1, C_2)} + \ldots \qquad 
\end{eqnarray}
The subscript ``twist" is used to remind the reader that a judiciously chosen boundary condition is used. 
Here, $m_g(L)$ is the ``mass gap for gauge fluctuations," which is unambiguously defined in the semiclassical, center-symmetric, small-$L$ regime.

We believe that for IR-CFTs with $N_f^{*} < N_f < N_f^{AF}$, the mass gap obtained in this way will exhibit   an $m(L) \sim \frac{1}{L} e^{-a S_0}$ scaling, where $a$ is a number depending on the details of the theory ($a=1$ for magnetic bions, which is the most generic case).  

For confining gauge theories with continuous global symmetries,  we expect:
\begin{eqnarray}
\langle W(C_1) W(C_2) \rangle|_{\rm twist} (L)  = \left\{ 
\begin{array}{ll}
 e^{-m_g(L) d(C_1, C_2)} + \ldots ~ &  LN\Lambda \ll1 \\
   ({\rm Free \; IR-limit}) + e^{-m_g(L) d(C_1, C_2)} + \ldots ~ &   LN\Lambda \gg 1~,
\end{array}
\right. 
\end{eqnarray}
where  $LN\Lambda \ll1$ is the domain of abelian confinement and $m_g(L)$ is expected to show the profiles  shown in Fig.~\ref{fig:massgap}{\bf a} or {\bf c}.  $LN\Lambda \gg1$ is the domain of non-abelian confinement, and it is expected that the mass gap in gauge sector will saturate to its $\R^4$ value (in particular, no $1/L$ type scaling should appear here and uniformity must set in).  
 In most theories of this class, these two regimes 
are split by a single chiral phase transition.  Nevertheless, we hope that our operator description  in terms of two-point connected correlators can be usefully applied to pin down the 
conformal window numerically.\footnote{This may be technically challenging: simulations  in sufficiently large volumes, using 
massless fermions with a (possibly chiral) twist (\ref{twist}), must be accessible in order to separate the different scales, e.g. $L^{-1}$$\ll$$m_g \sim \Lambda$$\ll$$1/a$, where $a$ is the lattice spacing. } 

Note that in IR-CFTs probed by such connected correlators, the path to large volume is smooth---the theory is always in the confinement without $\chi$SB regime with a  decreasing mass gap which vanishes in the decompactification limit.  
For confined theories, there is a critical radius where there is (typically) a chiral transition. The  absence or presence of this singular behavior is also sufficient to deduce whether the corresponding theory on $\R^4$ flows to conformality or confinement in the long-distance regime.

\end{document}